\documentclass[twocolumn,amsmath,amssymb,superscriptaddress]{revtex4-2}
\usepackage{graphics,amssymb,amsmath,epsfig,color,textgreek}
\usepackage{amsthm}
\usepackage{mathtools}
\usepackage{graphicx}
\usepackage[dvipsnames]{xcolor}
\usepackage{dcolumn}% Align table columns on decimal point
\usepackage{bm}% bold math
\usepackage[colorlinks=true,citecolor=cyan]{hyperref}
\hypersetup{colorlinks=true,citecolor=cyan,linkcolor=red,urlcolor=magenta}
\usepackage{braket}
\usepackage[normalem]{ulem}
\usepackage{cancel}
\usepackage{diagbox}

\usepackage{tikz}

\begin{document}

\newcommand{\beq}{\begin{equation}}
\newcommand{\eeq}{\end{equation}}
\newcommand{\bE}{{\bm E}}
\newcommand{\bT}{{\bm T}}
\newcommand{\bc}{{\bm c}}
\newcommand{\bk}{{\bm k}}
\newcommand{\br}{{\bm r}}
\newcommand{\bB}{{\bm B}}
\newcommand{\bH}{{\bm H}}
\newcommand{\bA}{{\bm A}}
\newcommand{\bh}{{\bm h}}
\newcommand{\bn}{{\bm n}}
\newcommand{\bp}{{\bm p}}
\newcommand{\bS}{{\bm S}}
\newcommand{\bP}{{\bm P}}
\newcommand{\bD}{{\bm D}}
\newcommand{\bx}{{\bm x}}
\newcommand{\by}{{\bm y}}
\newcommand{\bz}{{\bm z}}
\newcommand{\bu}{{\bm u}}
\newcommand{\bv}{{\bm v}}
\newcommand{\bw}{{\bm w}}
\newcommand{\bsigma}{{\bm \sigma}}
\newcommand{\btau}{{\bm \tau}}
\newcommand{\bchi}{{\bm \chi}}
\newcommand{\blambda}{{\bm \lambda}}
\newcommand{\bGamma}{\mathbf{\Gamma}}
\newcommand{\altmu}{\ensuremath{\text{\textmu}}}
\newcommand{\altnu}{\ensuremath{\text{\textnu}}}

\newcommand\norm[1]{\left\lVert#1\right\rVert}

\theoremstyle{definition}
\newtheorem{definition}{Definition}[section]

\newtheorem{theorem}{Theorem}[section]
\newtheorem{corollary}{Corollary}[theorem]
\newtheorem{lemma}[theorem]{Lemma}

\title{Signatures of Non-Abelian Kitaev quantum spin liquids in noise magnetormetry}

\author{Xiao Xiao}
\affiliation{Department of Physics, Northeastern University, Boston, Massachusetts 02115, USA}
\affiliation{Quantum Materials and Sensing Institute, Northeastern University, Burlington, Massachusetts 01803, USA}

\author{Masahiro O. Takahashi}
\affiliation{Department of Materials Engineering Science, Osaka University, Toyonaka 560-8531, Japan}

\author{Paul Stevenson}
\affiliation{Department of Physics, Northeastern University, Boston, Massachusetts 02115, USA}
\affiliation{Quantum Materials and Sensing Institute, Northeastern University, Burlington, Massachusetts 01803, USA}

\author{Satoshi Fujimoto}
\affiliation{Department of Materials Engineering Science, Osaka University, Toyonaka 560-8531, Japan}
\affiliation{Center for Quantum Information and Quantum Biology, Osaka University, Toyonaka 560-8531, Japan}

\author{Arun Bansil}
\affiliation{Department of Physics, Northeastern University, Boston, Massachusetts 02115, USA}
\affiliation{Quantum Materials and Sensing Institute, Northeastern University, Burlington, Massachusetts 01803, USA}

\date{\today}
\begin{abstract}
Identification of isolated Majorana zero modes (MZMs) is a key
step towards the realization of fault-tolerant topological quantum
computation. Here we show how the $T_1$-based noise magnetormetry
of a nitrogen-vacancy (NV) center qubit can reveal the unique
signatures of Majorana fermions attached to vacancies in a non-Abelian
Kitaev quantum spin liquid (KQSL). The $1/T_1$ of the NV center is found to be increased significantly when the working frequency of the NV center matches the energy difference between a MZM and a low-energy hybridized mode involving dangling Majorana fermions adjacent to vacancies. In experiments, this energy difference can be tuned by an external Zeeman field. Because of the large excitation gap of flipping a local $Z_2$ gauge field, the $1/T_1$ spectrum is robust against other fluctuations in KQSLs. Our study presents a promising pathway for identifying the non-Abelian phase in Kitaev materials.
\end{abstract}
\pacs{}
\maketitle

%\tableofcontents

%\section{Steps towards success}
%\input{procedures.tex}
%\section{Fixed point Hamiltonian}
%\input{fixed point}

{\color{cyan}{\textit{Introduction}}}. The celebrated Kitaev spin
model \cite{Kitaev} on a honeycomb lattice bridges the physics of
topological quantum matter and quantum information science. When the Kitaev interactions deviate from the so-called isotropic point, the ground state of the model becomes gapped \cite{Kitaev} and can be mapped to states within the code space of surface codes \cite{Kells}, which are used for quantum error correction \cite{Terhal}. Interestingly, near the isotropic point, the model transitions to a non-Abelian KQSL with the help of a time-reversal breaking Zeeman field. This exotic phase hosts MZMs bound to Ising anyons emergent as bulk excitations \cite{Kitaev}, which provide a foundation for fault-tolerant topological quantum computation \cite{Nayak}. Thus, Kitaev materials offer another pathway to fault-tolerant topological quantum computation other than topological superconducting materials \cite{Qi,Alicea,Sarma,Sato}.

Various materials have been identified as approximate realization of
the Kiteav honeycomb model
\cite{Jackeli,Chaloupka,Plumb,Kubota,Rousochatzakis,Sandilands,Kim,Kim16,Nasu,Arnab,
  Arnab17,Trebst,Takagi,Hermanns}. Among these, ruthenium trichloride
$\alpha$-RuCl$_3$ stands out as the most promising candidate. Recent
experiments suggest that a KQSL phase may emerge in $\alpha$-RuCl$_3$ under an in-plane magnetic field \cite{Baek,Sears,Wolter,Leahy,Jansa,Arnab18,Hentrich,Widmann,Balz,Czajka,
  Tanaka,Czajka23}, where a quantized thermal Hall conductance, attributed to a chiral Majorana edge state, has been reported \cite{Kasahara,Yamashita,Yokoi,Bruin}, although other studies found behaviors inconsistent with a KQSL phase within the same parameter regime \cite{Sahasrabudhe,Bachus}. In this connection, a variety of scanning tunneling setups have been examined theoretically with an eye towards unambiguous detection of fractionalized excitation characteristic of KQSLs \cite{Aasen,Konig,Pereira,Feldmeier,Udagawa,Bauer,Takahashi,Kao,Halasz}. However, little is available in literature beyond the tunneling spectroscopy, highlighting the clear need to explore alternative experimental approaches for probing KQSLs \cite{Go, Choi}.

\begin{figure}[h]
  \centering
  \includegraphics[width=1\columnwidth]{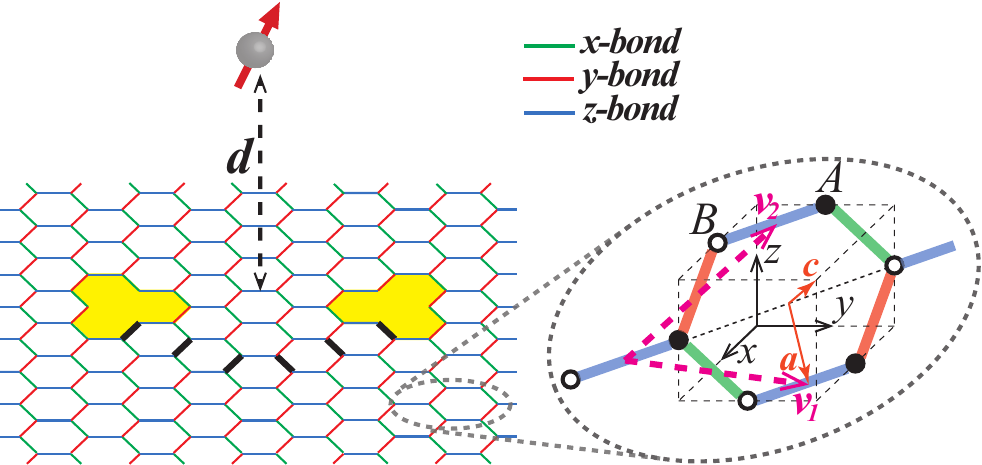}
  \caption{Schematic experimental setup showing an NV center qubit 			(gray ball with a red arrow) located at height $d$ above a Kiteav 		lattice 	with a pair of vacancies (yellow shaded regions) in the 			bound-flex sector, which is realized by flipping the $Z_2$ gauge 			fields on the black 	bonds along a string connecting the two 				vacancies. The zoom in shows an embedding of the Kitaev lattice 			into a $3D$ lattice of ligands in a real material, where the spin 		components are defined along the $x$, $y$ and $z$ axes, the $c$-axis 		is along $[111]$ direction perpendicular to the Kitaev lattice, and 		the $a$-axis is along $[11\bar{2}]$ direction within the 	Kitaev 			lattice plane. The unit cell consists of $A$ and $B$ sublattices 			connected by a $z$-bond. Magenta dashed arrows mark the two primitive 	vectors $\bv_1$ and $\bv_2$.}
   \label{fig1}
\end{figure}

In this letter, we focus on the bound-flux sector, which represents the ground state of site-diluted non-Abelian KQSLs \cite{Willans,Willans11,Kao20,Imamura24}. We show that the unique features of fractional excitations can be detected using $T_1$-based noise magnetometry with an NV center qubit \cite{Casola,Agarwal,Ariyaratne,Dolgirev,Khoo,Lee}, see Fig.~\ref{fig1}. In site-diluted KQSLs, external Zeeman fields can
release the localized dangling Majorana fermions adjacent to the vacancies and hybridize them with the originally itinerant Majorana in the system. This leads to the emergence of low-energy hybridized modes and MZMs trapped at vacancies as fractional excitations. Consequently, the wave functions of these modes exhibit significant overlap with the trapped MZMs at vacancy-adjacent sites, leading to strong spin correlations. The $T_1$ relaxation time of an NV center becomes sensitive to these spin correlations when its working frequency matches the energy difference between the hybridized modes and the MZMs. The external Zeeman field acts as a knob for adjusting aforementioned energy difference, enabling the NV center to detect signatures of the fractional excitations. Notably, because the large excitation gap must be overcome to flip a local $Z_2$ gauge field, we expect other spin fluctuations in the KQSLs to be suppressed at sufficiently low temperatures.

{\color{cyan}{\textit{The $T_1$ relaxation time of an NV center}}}. An NV center qubit located at a height $d$ above the Kitaev lattice can be described by the Hamiltonian $H_{qb}=\frac{\Omega}{2} \bn_q \cdot \bsigma$. Here $\Omega$ represents the working frequency of the qubit, and $\bn_q$ denotes the unit vector along the NV's quantization direction. The spin fluctuations in the Kitaev material couple to the NV center via magnetic-dipole interactions \cite{Chatterjee,Nieva,Langsjoen,support}, which effectively generates a magnetic field $\bB$ acting on the NV center: $H_c = g_q\mu_B \bB\cdot\bsigma$ with $g_q=2$ the $g$-factor of the NV center and $\mu_B$ the Bohr magneton.

The $T_1$ relaxation time of the NV center due to the fluctuating magnetic field can be obtained by the Fermi's golden rule \cite{Chatterjee,Nieva,Langsjoen}. At finite temperature, a straightforward calculation yields \cite{support}:
\beq \label{T1}
\frac{1}{T_1} = \sum_{i,j} \frac{\mathcal{I}_0 (1+e^{-\beta \Omega})}{r_i^3 r_j^3} \mathrm{Tr} \left[ \bchi_{i,j}(\Omega) \cdot U_{j;i}^{+;-} \right],
\eeq
where $r_j$ denotes the distance between the NV center and site $j$ on the Kitaev lattice, $\beta=1/k_B T$ denotes the inverse thermal energy, $\mathcal{I}_0 = g_q^2 g_s^2 \mu_B^4 \mu_0^2/(32\pi^2 \hbar)$ is a universal constant with $g_s$ denoting the $g$-factor of the Kitaev material and $\mu_0$ the vacuum permeability, the spin correlation matrix $\bchi_{i,j}(\Omega)$ has elements given by: $\bchi_{i,j}^{\mu,\nu}(\Omega) = \int dt e^{i\omega t} \langle S_i^\mu(t) S_j^\nu(0) \rangle$, and $U_{i;j}^{\mu;\nu} = [\bv_i^\mu]^T \cdot \bv_j^\nu$ is constructed from the orientation vectors $\bv_j^\nu$ in the the clockwise and anti-clockwise chiral directions with respect to the quantization direction $\bn_q$. The detailed expressions for $\bv_j^\nu$ are provided in \cite{support}.

{\color{cyan}{\textit{Model of site-diluted Non-Abelian KQSLs}}}. We
consider a site-diluted Non-Abelian KQSL described by the Hamiltonian of a Kitaev honeycomb model with vacancies at the isotropic point
\cite{Takahashi,Willans,Willans11,Kao20}:
\beq \label{Ham} H = -J
\sum_{i,j \notin \mathcal{V}, \langle i,j \rangle_{\alpha}}
S_{i}^{\alpha} S_{j}^{\alpha} - g_s \sum_{j \notin \mathcal{V}} \bh \cdot
\bS_j, \eeq
where $\langle i,j \rangle_{\alpha}$ refers to a bond connecting two neighboring sites $i$ and $j$, whose $\alpha$-component of the spins are coupled, and $\mathcal{V}$ represents the set of vacancy sites. The
time-reversal breaking Zeeman term in Eq.~(\ref{Ham}) drives the
system from the gapless phase to the non-Abelian phase
\cite{Kitaev}. Following Kitaev, we express the spins in terms of
Majorana fermions $S_j^\alpha = \frac{i}{2} b_j^\alpha c_j$. The
first term then becomes: $-J/4 \sum_{i,j \notin \mathcal{V},
  \langle i,j \rangle_{\alpha}} u_{\langle i,j \rangle_\alpha} c_i
c_j$, where $u_{\langle i,j \rangle_\alpha} = i b_i^\alpha b_j^\alpha =
\pm1$ denotes the conserved $Z_2$ gauge field associated with the
bond. We focus on the regime that $J$ is much larger than the Zeeman energy of a Kitaev spin $E_Z = |g_s \bh \cdot \bS_j|$ so that we can treat the Zeeman term perturbatively \cite{Kitaev}. For sites away
from the vacancies, the effective hopping of $c$-Majorana
between next-nearest neighbors is generated. In the contrast, the dangling $b$-Majorana adjacent to the vacancies can no longer condense to form a local $Z_2$ gauge field. Instead, they couple bilinearly to $c$-Majorana fermions in their vicinity with the help of the Zeeman coupling. In this sense, the Zeeman field `releases' these $b$-Majorana fermions. We define the itinerant Majorana $\psi_{\mathsf{j}}$ to include both the originally itinerant $c$-Majorana fermions $\psi_{\mathsf{j}=j}=c_j$ and the released $b$-Majorana fermions $\psi_{\mathsf{j}=j^\alpha}=b_j^\alpha$. Consequently, the Hamiltonian Eq.~(\ref{Ham}) becomes quadratic and can be expressed as $H \approx \sum_{\mathsf{j},\mathsf{k}} i \mathcal{T}(\{u_{\langle  \mathsf{j},\mathsf{k} \rangle}\}) \psi_{\mathsf{j}} \psi_{\mathsf{k}}$, where the effective hopping strength $\mathcal{T}(\{u_{\langle \mathsf{j},\mathsf{k} \rangle}\})$ depends on the $Z_2$ gauge field on the bonds connecting the sites $j$ and $k$ \cite{support}.

{\color{cyan}{\textit{Spin correlations in site-diluted Non-Abelian KQSLs}}}. The spin correlation at finite temperature is given by
\beq \label{spin_cor} \langle S_i^\mu(t) S_j^\nu(0) \rangle =
\frac{-1}{4} \frac{\mathrm{Tr}\left[ P e^{-(\beta-it) H} c_i b_i^\mu
    e^{-iHt} b_j^\nu c_j \right]}{\mathrm{Tr}\left[ Pe^{-\beta
      H}\right] }, \eeq
where $P$ projects the wave function onto the physical subspace.

When $i,j \notin \mathcal{N}$, a set of sites adjacent to the vacancies, the $b$-Majorana involved in Eq.~(\ref{spin_cor}) are localized. One can prove that $b_i^\mu e^{-iHt} = e^{-i\tilde{H}t} b_i^{\mu}$, where the Hamiltonian $\tilde{H}$ is obtained from $H$ by flipping the sign of the $Z_2$ gauge field on the $\mu$ bond connecting to the site $i$ \cite{Baskaran}. At sufficiently low temperatures, the spin correlation can be evaluated using the ground state wave function as: $\langle S_i^\mu(t) S_j^\nu(0) \rangle = \langle \mathcal{M}| e^{itH} c_i e^{-i\tilde{H}t} c_j |\mathcal{M}\rangle \langle \mathcal{G}| b_i^\mu b_j^\nu
|\mathcal{G}\rangle$, where $|\mathcal{M}\rangle$ and $|\mathcal{G}\rangle$ denote the itinerant matter and the gauge field sectors of the ground state. This result shows that the spin correlation vanishes unless $b_j^\nu = b_i^\mu$ or $b_i^\mu b_j^\nu =-i u_{\langle i,j \rangle_\mu} \delta_{\mu,\nu}$. This selection rule is well known for the Kiteav model and reflects the short-range nature of the spin correlations. In the Lehmann representation the spin correlation is written as:
\beq \label{spin_away} \langle S_i^\mu(t) S_j^\nu(0) \rangle =
\frac{-1}{4} \sum_{\lambda} e^{-it(E_\lambda-E_0)} \langle
\mathcal{M}| c_i |\lambda \rangle \langle \lambda| c_j
|\mathcal{M}\rangle, \eeq
where $|\lambda\rangle$ denotes the eigenstates of $\tilde{H}$ with
eigenenergies $E_\lambda$. Let us denote the many-body ground state of $\tilde{H}$ as $|\lambda_0\rangle$ so that all others $|\lambda \rangle$ can be constructed from $|\lambda_0\rangle$ by creating excitations. We find that $\Delta_\lambda = E_\lambda-E_0$ has the minimal value $\Delta_0 = E_{\lambda_0}-E_0$, which is known as the flux gap. Equation~(\ref{spin_away}) indicates that the spin correlation vanishes for frequencies below the flux gap $\Delta_0$ \cite{Knolle}.

When $i,j \in \mathcal{N}$, the $b$-Majorana in Eq.~(\ref{spin_cor}) are itinerant and thus included in the Hamiltonian $H$. In this case, one can show that $\psi_{\mathsf{j}} e^{-iHt} = \sum_{\mathsf{k}} [e^{-\mathcal{T}t}]_{\mathsf{k},\mathsf{j}} e^{-iHt} \psi_{\mathsf{k}}$, where
$\mathcal{T}$ denotes the hopping matrix. Using Wick's theorem, the spin correlation can be expressed in the following informative form:
\beq \label{spin_vac} \langle S_i^\mu(t) S_j^\nu(0)
\rangle = \frac{-1}{4} \sum_{n,m} \mathcal{C}(\{E_n\},\beta)
e^{-i\mathcal{E}_{n,m}t} F^{\mu,\nu}(E_n,E_m), \eeq
where: $\mathcal{C}(\{E_n\},\beta)$
depends on the spectrum of the Hamiltonian $\{E_n\}$ and the
temperature, $\mathcal{E}_{n,m}=E_n+E_m$, and $F^{\mu,\nu}(E_n,E_m)$ is determined by the eigenfunctions for $E_n$ and $E_m$ at the vacancy-neighboring sites (see \cite{support} for details).

\begin{figure}[h!]
  \centering
  \includegraphics[width=1\columnwidth]{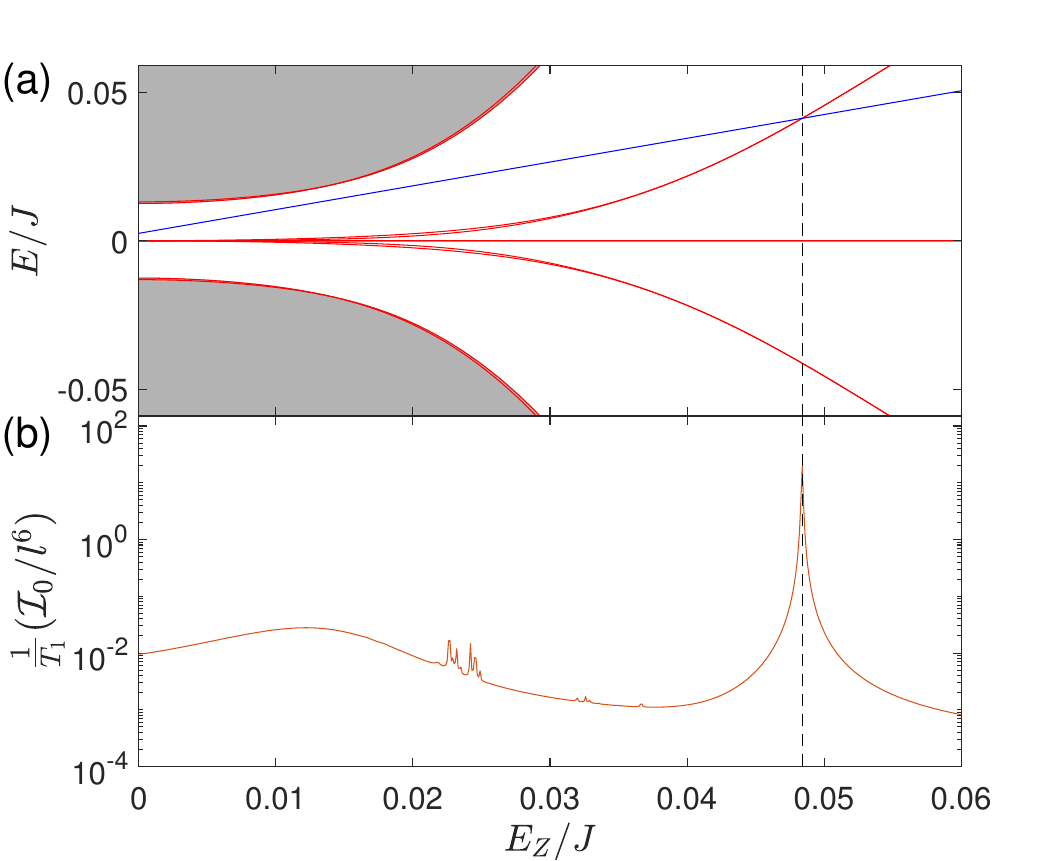}
  \caption{Non-Abelian KQSL with a pair of vacancies in the bound-flux
    	sector: (a) the low-energy portion of the spectrum as a function 			of Zeeman energy of a Kitaev spin $E_Z$, with hybridized modes 			having significant wave function weights on vacancy-neighboring 			sites highlighted in red. For comparisons, the working frequency 			of an NV center is indicated by the blue line. (b) The $1/T_1$ 			spectrum of an NV center at $d=l=30a$ ($a$ denoting the lattice 			constant) above the center of the calculation supercell shows 			significant $1/T_1$ enhancement when the energy differences 				between hybridized modes and MZMs matches the working frequency 			(indicated by the vertical dash line). Negligibly small 					fluctuations due to bulk modes are also observed.}
   \label{fig2}
\end{figure}

{\color{cyan}{\textit{$T_1$-based noise magnetometry in Non-Abelian KQSL}}}. Now we are in a position to qualitatively understand $T_1$-based noise magnetometry for site-diluted non-Abelian KQSLs. Typically, the working frequency of an NV center at zero field is $\sim 3$~GHz ($\Omega_0 \sim 0.013$~meV) \cite{Casola}, and the Kitaev coupling strength $J$ is in the range $2.8-6.8$meV \cite{Eichstaedt}. In the calculations below we set $\Omega_0/J=1/250$. In the presence of a Zeeman field, the working frequency of an NV center relates to the Zeeman energy of a Kitaev spin as $\Omega = \Omega_0+g_q/g_s E_Z$. The flux gap, below which the spin correlation for spins away from the vacancies vanishes, is $\sim 0.065J$ \cite{Knolle}. This value can be larger than $\Omega$ in the small Zeeman field regime, rendering their contributions negligible. The contribution of spins neighboring the vacancies depends on the energy spectrum of $H$. When the Zeeman coupling is turned on, the $b$-Majorana on these sites are released and hybridize with one another itinerant Majorana (including both $c$-Majorana and themselves) to form finite but extremely low energy states. Because these states arise from the hybridization of $b$-Majorana originally locating on sites adjacent to the vacancies, the associated wave functions are expected to display significant overlap at these sites, leading to large values of $F^{\mu,\nu}(E_n,E_m)$. The Zeeman field acts as a tuning parameter, allowing the spectrum of these hybridized modes to be adjusted to match the working frequency $\Omega$. When this condition is met, a significant $1/T_1$ enhancement is expected.

\begin{figure}[h!]
  \centering
  \includegraphics[width=1\columnwidth]{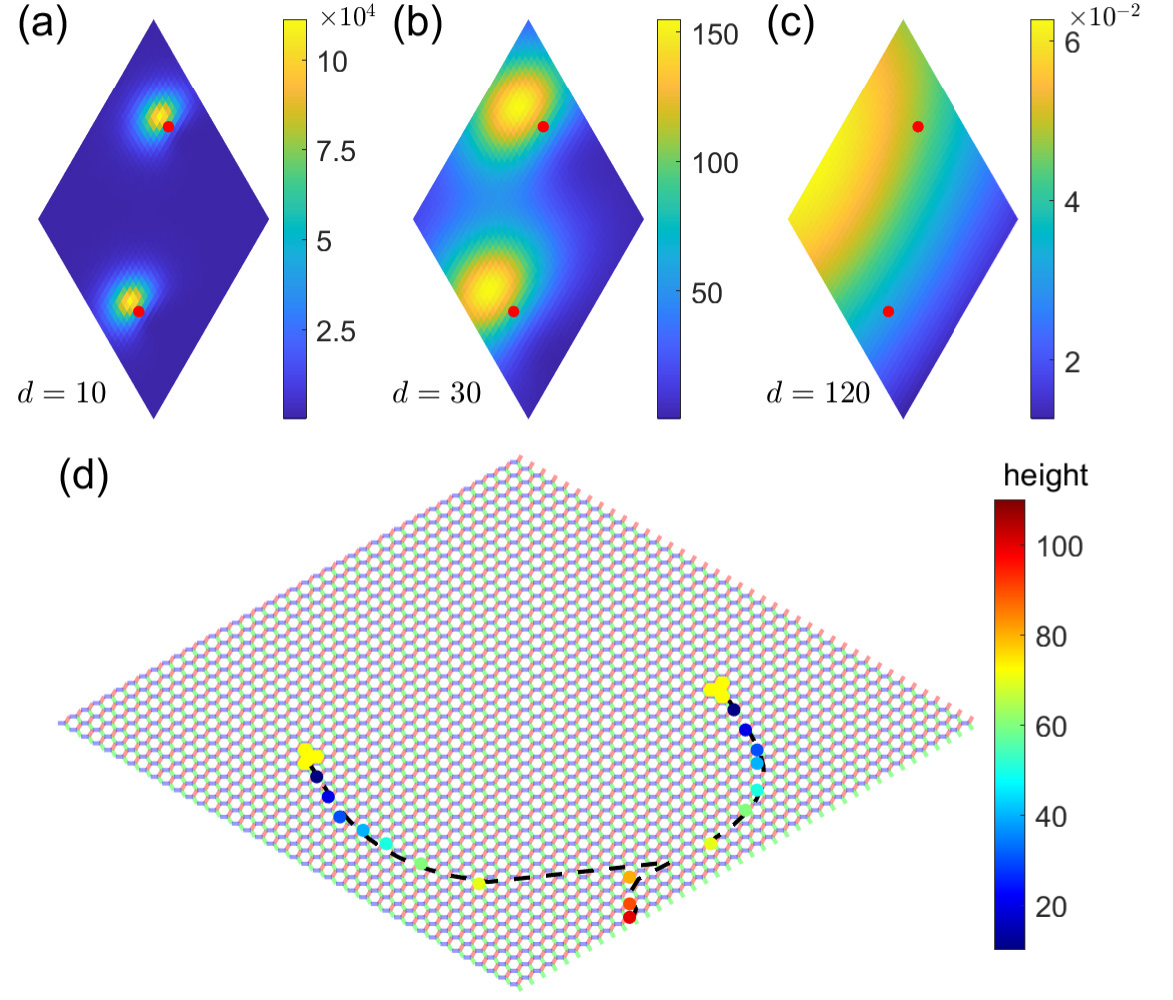}
  \caption{Spatial property of $1/T_1$ on the Kitaev lattice: (a),
    (b), and (c) show the $1/T_1$ spatial distributions for NV
    centers at heights $d=10a$, $d=30a$, and $d=120a$,
    respectively. The two red dots represent the positions of the
    vacancies. (d) illustrates the evolution of the locations of maximal 		$1/T_1$ as the NV center height $d$ changes. The black dashed
    curves are obtained from the approximation Eq.~(\ref{T1_app}), while
    the colored dot, according to different NV center heights, are 			obtained rom Eq.~(\ref{T1}). All results use the same parameters as    	in Fig.~\ref{fig2} at the resonance.}
   \label{fig3}
\end{figure}

{\color{cyan}{\textit{Results and discussion}}}. To support the qualitative understanding, we focus on the minimal case of a pair of vacancies in the bound-flux sector on a $40 \times 40$ unit cell lattice, indexed by $(n,m)$ along the two primitive directions (Fig.~\ref{fig1}). Vacancies are placed at the $A$-sublattice of the unit cell $(32,27)$ and the $B$-sublattice of the unit cell $(9,14)$, with the NV center polarization assumed to be along the $x$-direction and the Zeeman field aligned along the $a$-axis.

When the field is applied, the originally localized $b$-Majorana adjacent to the vacancies are released, allowing them to hybridize with other itinerant Majorana into new modes. As shown in Fig.~\ref{fig2}(a), two MZMs are trapped at the vacancies, while other hybridized modes, with significant wave function weights on the vacancy-neighboring sites, gain finite energies. As the Zeeman field varies, the energy differences between hybridized modes and MZMs widen, and eventually some of them match $\Omega$ (the blue line in Fig.~\ref{fig2}(a)). Under this condition, a significant enhancement of $1/T_1$ is shown in Fig.~\ref{fig2}(b), indicating that spin fluctuations mediated by MZMs and the hybridized modes are manifested as the $1/T_1$ enhancement. 

In Figs.~\ref{fig3}(a), (b), and (c), the spatial distributions of
$1/T_1$ are shown on the Kiteav lattice for three different heights of the NV centers. The maximal enhancement of $1/T_1$ is not observed at the locations of the vacancies. This behavior is distinct from the case of an Anderson impurity \cite{Agarwal} and highlights the correlated nature of the hybridized modes. To understand it, we start from Eq.~(\ref{T1}) with the observation that only sites adjacent to the vacancies contribute. Given that the distances between the NV center and vacancies are much larger than $a$, we can approximate the positions of the spins by those of their adjacent vacancies. Due to the locality of the wave functions, the correlation between spins adjacent to different vacancies can be ignored. Consequently, $1/T_1$ can be expressed in terms of the spin correlation matrix $\bchi_{\Lambda,\Lambda}(\Omega)$ and the orientation matrix $U_{\Lambda;\Lambda}^{+;-}$ for the $\Lambda$ vacancy. We find that both Hermitian matrices have only one principal component. Therefore, $1/T_1$ can be approximated as \cite{support}
\beq \label{T1_app} \frac{1}{T_1} \approx
\mathcal{I}_0 (1+e^{-\beta\Omega}) \sum_{\Lambda} \frac{\xi_{\Lambda}
  \xi_{U_\Lambda}}{r_\Lambda^6} \left| \bv_{\xi_{U_\Lambda}} \cdot
\bv_{\xi_\Lambda} \right|^2, \eeq
where $\xi_\Lambda$ is the principal value of $\bchi_{\Lambda,\Lambda}(\Omega)$ with its corresponding principal vector $\bv_{\xi_\Lambda}$, and $\xi_{U_{\Lambda}}$ is the principal value of $U_{\Lambda;\Lambda}^{+;-}$ with its principal vector $\bv_{\xi_{U_\Lambda}}$. Eq.~(\ref{T1_app}) suggests that the spatial dependence of $1/T_1$ is determined by two factors: (i) the distance between vacancies and the NV center; and (ii) the alignment of the Majorana wave function polarization at a vacancy with the spatial orientation of the vector connecting the vacancy and the NV center. The second factor causes the maximum of $1/T_1$ to deviate from the positions of the vacancies. In Fig.~\ref{fig3}(b), we compare the trace of local maximum in $1/T_1$ as it changes with the NV center height. Results obtained from Eq.~(\ref{T1_app}), shown as black dashed curves, are consistent with those obtained from Eq.~(\ref{T1}), shown as colored dots according to the heights. The good agreement validates the two factors aforementioned.

Regarding experimental considerations, we estimate the properties of $1/T_1$ spectrum based on the results shown in Figs.~\ref{fig2} and \ref{fig3}. At temperatures of $\sim$1K and Zeeman fields $\sim$1T, we find that for NV centers at a height $d=30a\sim 10$nm  over $1/3$ of the supercell can have an $1/T_1$ increment larger than $30$Hz, while for NV centers at a height $d=10a\sim 3.5$nm over $1/3$ of this rises to $3$kHz \cite{support}. Recent experiments have demonstrated milisecond $T_1$ times for shallow ($<10$nm) NV centers \cite{Myers,Sangtawesin}, which is sufficient to detect $1/T_1$ enhancement on the order of $100$Hz. Low temperature will further improve the sensitivity of NV centers \cite{Jarmola}. A major challenge in experiments is placing NV centers close enough to the maximal enhancement positions. The fact that only $1/3$ of each supercell has a relaxation rate $>20$Hz at a separation $10$nm, further compounded by non-even distribution of vacancies in real materials, makes the lateral localization challenging. This requires the measurement of many NV centers to find one optimally aligned, though the distinct magnetic field dependence (Fig.~\ref{fig2}) will aid in identifications. On the materials side, however, the advancement of on-demand generation of defects will aid significantly \cite{Roccapriore,Thomas}. We note that many prospective KQSL are $2D$ materials which can be integrated directly onto the diamond surface by exfoliation \cite{Zhang,Zhou,Kumar}. Finally the high field and low temperature detecting environment need also to be addressed. Large fields require careful alignment of the external field with the NV center quantization axis to avoid quenching the optical detection of the spin state \cite{Welter,Beaver}. Room temperature operation at high fields has been demonstrated previously \cite{Stepanov,Fortman,Ren}. While extending these approaches to the low temperature regime is a significant feat of engineering, it is close to the capabilities offered by many commercial instruments, giving us confidence that this is not an insurmountable obstacle. 

{\color{cyan}{\textit{Summary}}}. We discussed how $T_1$-based
noise magnetometry can provide a new pathway for probing the non-trivial correlations of low-energy modes arising from the released dangling Majorana fermions adjacent to vacancies as robust signatures of non-Abelian KQSLs. Notably, the experimental setup considered here might be also applied to other defects binding Ising anyons in KQSLs \cite{Takahashib}. Our study opens the door to use noisy magnetometry to explore fractionalized excitations in exotic quantum phases. 

{\color{cyan}{\textit{Acknowledgment}}}. We are grateful for the valuable discussions with Patrick Lee and Eduardo Mucciolo. The work of X.X. and A.B. was supported by the National Science Foundation through the Expand-QISE award NSF-OMA-2329067 and benefited from the resources of Northeastern University’s Advanced Scientific Computation Center, the Discovery Cluster, the Massachusetts Technology Collaborative award MTC-22032, and the Quantum Materials and Sensing Institute. M.O.T and S.F. are supported by JST CREST Grant No. JPMJCR19T5, as well as JSPS KAKENHI No. JP22J20066 and No. JP23K20828. M.O.T. also acknowledges the support from the Japan Society for the Promotion of Science (JSPS) Fellowship for Young Scientists.

\pagebreak
\widetext

\setcounter{figure}{0}
\renewcommand{\thefigure}{S\arabic{figure}}
\renewcommand{\bibnumfmt}[1]{[S#1]}
\renewcommand{\citenumfont}[1]{S#1}
\setcounter{equation}{0}
\renewcommand{\theequation}{S\arabic{equation}}

\begin{center}
\textbf{\large Supplementary materials for: `Signatures of Non-Abelian Kitaev quantum spin liquids in noise magnetometry'}
\end{center}

In this document, we provide derivations of key formulas and details of the estimations presented in the main text. Additionally, we include more data to illustrate how the $1/T_1$ noise spectrum is influenced by the polarization of NV centers, the working temperature, and the orientation of the Zeeman field.

\tableofcontents

\section{The expression of $1/T_1$: Derivation of Eq.~(1)}

Without lose generality, the magnetic qubit is described by:
\beq
H_{qb} = \frac{\Omega}{2} \bn_{q} \cdot \bsigma,
\eeq
where $\bn_q=\bn'_q+\bB_0$ denotes the unit vector of total polarization direction of the qubit, with $\bn'_q$ representing the vector along the intrinsic spin polarization direction, and $\bB_0$ for the vector along the direction of external Zeeman field. Now we consider the influence of a fluctuation magnetic field generated by the nearby material, which couples to the qubit via:
\beq \label{qubit_couple}
H_{c} = g_q \mu_B \bB \cdot \bsigma,
\eeq
where $g_q$ is the $g$-factor of the qubit and $\mu_B$ is Bohr magneton. We denote the direction of $\bn_q$ as the $w$-direction, from which we form a orthogonal $\bu-\bv-\bw$ coordinate system. We denote the eigenstate of $\sigma_{w}$ with eigenvalue $-1$ as $|-\rangle$ and the eigenstate of $\sigma_w$ with eigenvalue $1$ as $|+\rangle$. Then the emission and absorption rate can be obtained through Fermi's golden rule:
\beq
\frac{1}{T_{abs, em}} = 2\pi \sum_{m,n} p_n \langle n| B^{\pm} |m\rangle \langle m | B^{\mp} |n\rangle \delta(\omega \pm E_{mn}),
\eeq
where $p_n$ denotes the probability of the state $|n\rangle$, $B^{\pm} = B^u\pm i B^v$ and $E_{mn}=E_m-E_n$ denotes the energy difference between the perturbed state $m$ and the initial state of the material $n$. Converted into integral form, the relaxation rate of the qubit is defined as the average of the absorption and the emission rates and can be expressed as:
\beq 
\frac{1}{T_1} = \frac{g_q^2 \mu_B^2}{2} \int_{-\infty}^{\infty} dt e^{i\Omega t} \langle [B^{-}(t),B^{+}(0)]_+ \rangle.
\eeq
One can show that the expression for $1/T_1$ can be differently as:
\beq \label{T1_general}
\frac{1}{T_1} = \frac{g_q^2 \mu_B^2}{2} (1+e^{-\beta \Omega}) \int_{-\infty}^{\infty} dt e^{i\Omega t} \langle B^{-}(t) B^{+}(0) \rangle
\eeq
which is equivalent to the form obtained from fluctuation dissipation theorem:
\beq 
\frac{1}{T_1} = g_q^2 \mu_B^2 \coth \frac{\beta \Omega}{2} \Im m \left[ \int_{0}^{\infty} dt e^{i\Omega t} \left\langle \left[ B^-(t), B^+(0) \right] \right\rangle \right].
\eeq

To derive Eq.~(1) in the main text, we need to determine the expression for the magnetic field generated by the fluctuations of the spins on the lattice. We can expect that the leading order coupling is the magnetic-dipole coupling between a spin and the qubit, whose Hamiltonian is given by:
\beq \label{dipole_couple}
H = - \frac{\mu_0 \gamma_q \gamma_s \hbar^2}{4\pi r^3} \left[ 3\left( \bS \cdot \hat{\br} \right) \left( \bsigma \cdot \hat{\br} \right) - \bS \cdot \bsigma \right],
\eeq
where $\mu_0$ is the permeability, $\gamma_s$ is the gyromagnetic ratio for the spin, $\gamma_q$ is the gyromagnetic ratio for the qubit, $\bS$ denotes the spin operator, $\br$ is the vector pointing to the qubit from the spin, $\hat{\br}$ is the unit vector along $\br$, and $r$ is the length of $\br$. The gyromagnetic ratio is given by:
\beq
\gamma_{s/q} = \frac{g_{s/q} \mu_B}{\hbar}.
\eeq
where $g_s$ denotes the $g$-factor for the spin and $g_q$ denotes the $g$-factor for the qubit. Casting Eq.~(\ref{dipole_couple}) into the form of Eq.~(\ref{qubit_couple}), we can find the expression of the fluctuation magnetic field generated on the qubit as:
\beq \label{Bfield}
\bB = \frac{g_s \mu_0 \mu_B}{4\pi} \sum_{j} \left[ \frac{\bS_j }{r_j^3} - \frac{3 \left(\bS_j \cdot \br_j\right) \br_j}{r_j^5} \right].
\eeq
We now need to insert this expression into Eq.~(\ref{T1_general}) to find the expression for $1/T_1$. We can write Eq.~(\ref{Bfield}) in components:
\beq
\begin{cases}
B^x = \frac{g_s \mu_0 \mu_B}{4\pi} \sum_j \frac{1}{r_j^3} \left[ \left( 1 - 3\tilde{x}_j^2 \right) S_j^x - 3\tilde{x}_j \tilde{y}_j S_j^y - 3 \tilde{x} \tilde{z}_j S_j^z \right], \\
B^y = \frac{g_s \mu_0 \mu_B}{4\pi} \sum_j \frac{1}{r_j^3} \left[ - 3\tilde{y} \tilde{x}_j  S_j^x + \left(1 -3\tilde{y}_j^2\right) S_j^y - 3 \tilde{y} \tilde{z}_j S_j^z \right], \\
B^z = \frac{g_s \mu_0 \mu_B}{4\pi} \sum_j \frac{1}{r_j^3} \left[ - 3\tilde{z} \tilde{x}_j  S_j^x - 3\tilde{z}_j\tilde{y}_j S_j^y + \left(1 - 3 \tilde{z}_j^2 \right) S_j^z \right].
\end{cases}
\eeq
This form motivate us to define the orientation vectors:
\beq \label{or_vec}
\begin{cases}
\bv_j^x = \left(\begin{array}{ccc} 1-3\tilde{x}_j^2 & -3 \tilde{x}_j \tilde{y}_j & -3 \tilde{x}_j \tilde{z}_j \end{array}\right),~ \\
\bv_j^y = \left(\begin{array}{ccc}  -3 \tilde{y}_j \tilde{x}_j & 1-3\tilde{y}_j^2 & -3 \tilde{y}_j \tilde{z}_j \end{array}\right), \\
\bv_j^z = \left(\begin{array}{ccc}  -3 \tilde{z}_j \tilde{x}_j & -3 \tilde{z}_j \tilde{y}_j & 1-3\tilde{z}_j^2 \end{array}\right),
\end{cases}
\eeq
with $\tilde{\alpha}_j = \br_j \cdot \hat{\alpha}/r_j$ for $\alpha=x,y,z$. Using them, we have:
\beq \label{compact_B}
B^\alpha = \frac{g_s \mu_0 \mu_B}{4\pi} \sum_j \bv_j^\alpha \cdot \bS_j.
\eeq
One can show generically that, for an arbitrary polarization of the qubit, the associated chiral vectors can be always constructed from the three basis vector defined in Eq.~(\ref{or_vec}). For example, if the qubit polarization is along the $x$-direction, the two clockwise ($\bv_j^+$) and anti-clockwise ($\bv_j^-$) chiral vectors with respect to the polarization direction $x$ are given by:
\beq
\bv_j^{\pm} = \bv_j^y \pm i \bv_j^z.
\eeq

Straightforwardly inserting Eq.~(\ref{compact_B}) into Eq.~(\ref{T1_general}), we obtain:
\beq 
\frac{1}{T_1} = \frac{g^2 g_s^2 \mu_0^2 \mu_B^4}{32\pi^2} (1+e^{-\beta \Omega}) \sum_{i,j} \frac{1}{r_i^3 r_j^3} \left[ \bv_i^{-} \cdot \left( \int dt e^{i\Omega t} \langle \bS_i(t) \bS_j^T(0) \rangle \right) \cdot \left(\bv_j^+\right)^T \right].
\eeq
We can define the orientation matrix from the chiral vectors: $U_{j;i}^{+;-} = (\bv_j^+)^T \bv_i^-$, and write in the trace form:
\beq \label{T1_main}
\frac{1}{T_1} = \frac{g^2 g_s^2 \mu_0^2 \mu_B^4}{32\pi^2} (1+e^{-\beta \Omega}) \sum_{i,j} \frac{1}{r_i^3 r_j^3} \mathrm{Tr} \left[  \bchi_{i,j}(\Omega) \cdot U_{j;i}^{+;-} \right],
\eeq
where the elements of $\bchi_{i,j}(\Omega)$ is given by:
\beq
\chi_{i,j}^{\alpha,\beta}(\Omega) = \int dt e^{i\Omega t} \langle S_i^\alpha(t) S_j^\beta(0) \rangle.
\eeq
Eq.~(\ref{T1_main}) is Eq.~(1) in the main text.

\section{Low-energy effective Hamiltonian of site-diluted Kitaev spin liquids}

We begin from the Hamiltonian of the site-diluted Kiatev spin liquid, Eq.~(2) of the main text:
\beq
H = -J \sum_{\alpha} \sum_{\langle j,k \rangle} S_j^\alpha S_k^\alpha - g_s \sum_{j} \bh \cdot \bS.
\eeq
For the simplicity, in the following derivation, we will absorb $g_s$ into the magnetic field and redefine $\bh = g_s \bh$. Following Kiteav \cite{KitaevS}, we can express the spin operator in terms of Majorana, which doubles the Hilbert space and can be regulated by a projection operator $P$. In particular, we have: $S_j^\alpha = \frac{i}{2} b_j^\alpha c_j$. The first term in Eq.~(1) can then be expressed in terms of Majorana fermions:
\beq
H_1 = \frac{-J}{4} \sum_\alpha \sum_{\langle j,k \rangle} \left( b_j^\alpha b_k^\alpha \right) c_j c_k.
\eeq
For the lattice that does not involve vacancies, we can define a local $Z_2$ gauge field: $u_{j,k}^\alpha = ib_j^\alpha b_k^\alpha = \pm1$, which pairs the two $b$-Majorana fermions on the same bond. Thus, the first term can be written as:
\beq \label{NNhopping}
H_1 = \frac{i}{4}\sum_\alpha \sum_{\langle j,k \rangle} J u_{j,k}^\alpha c_j c_k.
\eeq
This Hamiltonian suggests that $c$-Majorana fermions effectively hop from site to site and thus are itinerant. On the other hand, $b$-Majorana fermions are localized at the bond and condensed to form a local $Z_2$ gauge field. As shown by Kitaev \cite{KitaevS}, according to Lieb's theorem, the many-body ground state of the Hamiltonian $H_1$ should have uniformed $Z_2$ gauge field. We can choose them to be either $1$ or $-1$. Flipping a $Z_2$ gauge field generates an excited state by costing energy.

\begin{figure}[h!]
  \centering
  \includegraphics[width=0.5\columnwidth]{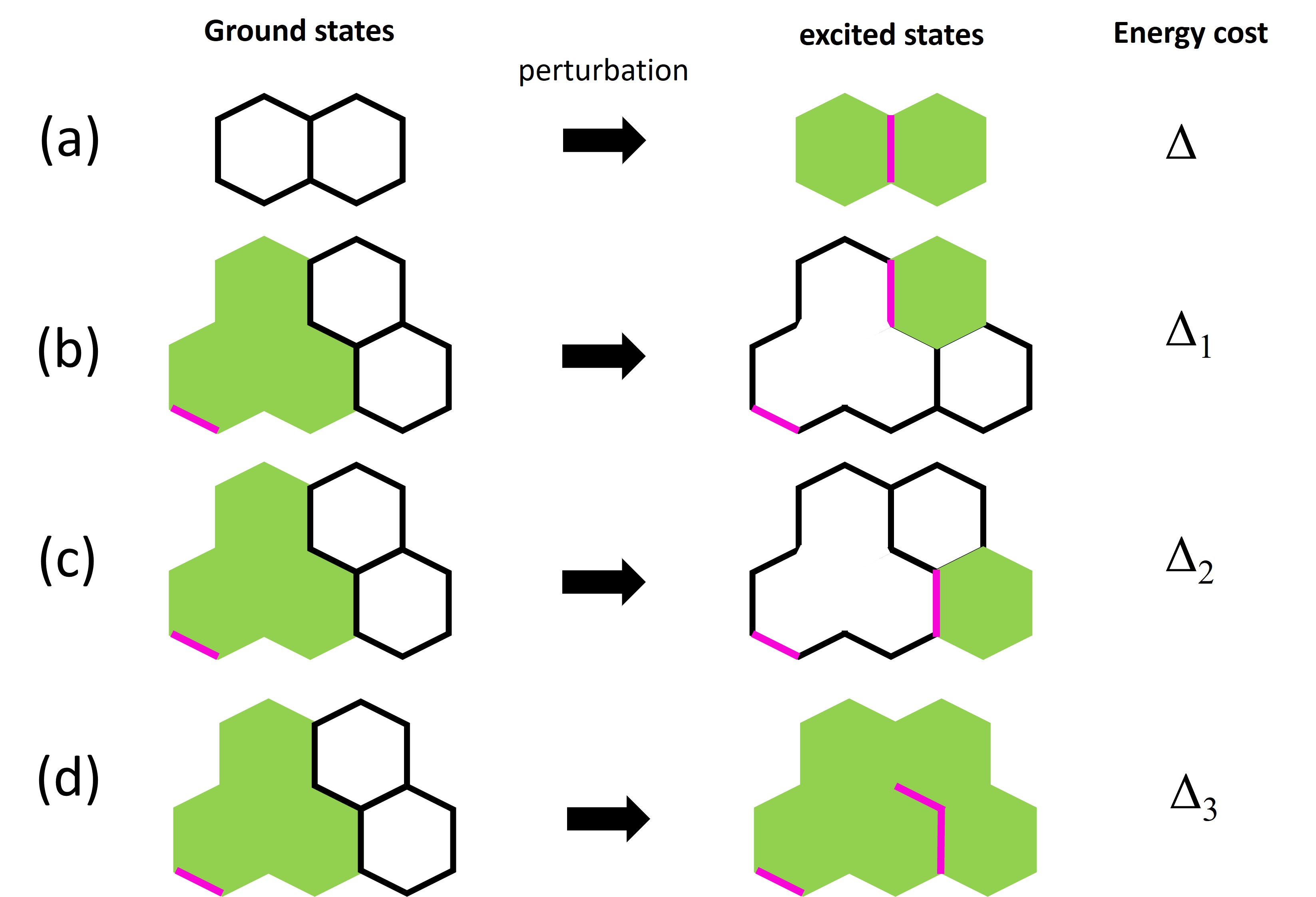}
  \caption{Creating excited states by flipping local $Z_2$ gauge fields: (a) the situation away from vacancies; (b), (c) and (d) show three different processes around a vacancy with different energy costs. The perturbations are represented by extra magenta bonds. The green colored regions attract one $Z_2$ flux.} 
   \label{figs1}
\end{figure}

\subsection{The effect of Zeeman field on $c$-Majorana fermions away from vacancies}

In general, the Zeeman term makes this model not exactly solvable. However, when the strength of Zeeman energy of a Kitaev spin is much smaller than the Kitaev coupling, namely $E_Z \ll J$, we can treat the Zeeman term perturbatively. Starting from the ground state, the application of a Zeeman term $h^\alpha S_j^\alpha=ih^\alpha b_j^\alpha c_j$ would flip the $Z_2$ field on the $\alpha$-bond connecting to the site $j$ due to $b_j^\alpha$. An intermediate perturbative state, which has an energy $\Delta$ higher, is thus generated. This process is shown in Fig.~\ref{figs1}(a). To return to the ground state, we need to acting a Zeeman term to the same bond. Consequently, the non-trivial leading order perturbation is given by:
\beq
H_{c,1}^{(1)} = -\sum_\alpha \sum_{\langle j,k \rangle} \frac{2(h^\alpha)^2}{\Delta} S_j^\alpha S_k^\alpha = \frac{i}{4} \sum_\alpha \sum_{\langle j,k \rangle} \frac{2(h^\alpha)^2}{\Delta} u_{j,k}^\alpha c_j c_k,
\eeq
whose effect is renormalized into the Kitaev coupling strength $J$. We can combine it with Eq.~(\ref{NNhopping}) to describe the effective hopping of $c$-Majorana fermions between the nearest neighboring sites:
\beq
H'_1 = \frac{i}{4}\sum_\alpha \sum_{\langle j,k \rangle} \tilde{J}_\alpha u_{j,k}^\alpha c_j c_k,
\eeq
where the renormalized hopping strength is $\tilde{J}_\alpha = J + \frac{2(h^\alpha)^2}{\Delta}$. It indicates that the original isotropic hopping becomes anisotropic through a polarized Zeeman field.

The next order non-trivial contribution comes from applying a Zeeman term to the three bonds connecting to the same vertex. One can easily verify that after transitions to two intermediate states with the same energy, the ground state is recovered. The order in which Zeeman terms act on the three bonds is independent, so the effective Hamiltonian at this order is given by:
\beq \label{2nd_order_away}
H_{c,2}^{(2)} = -\sum_{\langle j,k,l \rangle} \frac{6h^\alpha h^\beta h^\gamma}{\Delta^2} S_j^\alpha S_k^\gamma S_l^\beta = \frac{i}{4} \sum_{\langle j,k,l \rangle} \epsilon_{\alpha,\beta,\gamma} \kappa u_{j,k}^\alpha u_{l,k}^\beta c_j c_l,
\eeq
where $\kappa = \frac{6h^\alpha h^\beta h^\gamma}{2\Delta^2}$. Eq.~(\ref{2nd_order_away}) indicates that through the application of Zeeman field, an effective hopping of $c$-Majorana fermions on the sites, which are next-nearest-neighboring to each other, is generated.

\subsection{The effect of Zeeman term on the $c$-Majorana fermions in the vicinity of vacancies}

\begin{figure}[h!]
  \centering
  \includegraphics[width=0.5\columnwidth]{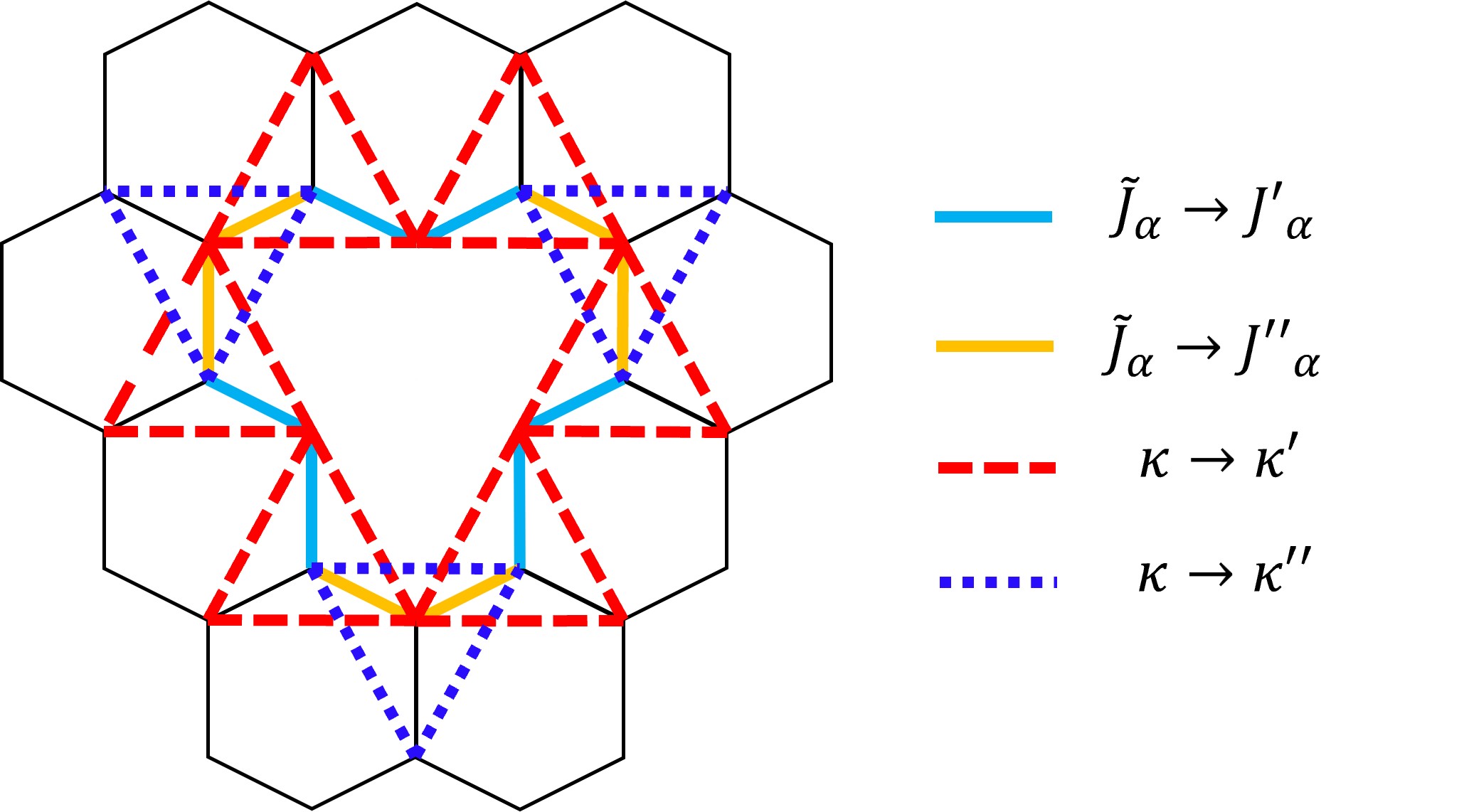}
  \caption{Illustration of the bonds on which the effective hopping strength of $c$-Majorana fermions changes due to the perturbation involving new excited states shown in Fig.~\ref{figs1}. The modifications to the parameters are specified in the text.} 
   \label{figs2}
\end{figure}

The influence of the Zeeman field on the $c$-Majorana fermions in the vicinity of vacancies follows the same perturbation strategy. The main difference is that flipping local $Z_2$ gauge fields around a vacancy leads to different excitation states. The corresponding processes and the resulting excited states are summarized in Fig.~\ref{figs1}(b), (c) and (d). These differences eventually modify the effective hopping parameters $\tilde{J}$ and $\kappa$ on the bonds near the vacancies. We summarize these changes around a vacancies explicitly in Fig.~\ref{figs2}.

In particular, for the nearest neighboring hopping through the bonds in cyan the hopping strength changes as:
\beq
\tilde{J}_\alpha\to \tilde{J}'_\alpha = \left( J + \frac{(h^\alpha)^2}{\Delta_1} \right).
\eeq
for the nearest neighbor hopping through the bonds in orange, the hopping strength changes as:
\beq
\tilde{J}_\alpha\to \tilde{J}''_\alpha =  \left( J + \frac{(h^\alpha)^2}{\Delta_2} \right).
\eeq
for the next-nearest neighboring hopping through the red dashed lines, the hopping strength changes as:
\beq
\kappa \to \kappa' = 2h^x h^y h^z \left( \frac{1}{\Delta_1 \Delta_2} + \frac{1}{\Delta_2 \Delta_3} + \frac{1}{\Delta_1 \Delta_3} \right),
\eeq
and, for the next-nearest neighboring hopping through the blue dotted lines, the hopping strength changes as:
\beq
\kappa \to \kappa'' = 2h^x h^y h^z \left( \frac{1}{\Delta_2^2} + \frac{2}{\Delta_2 \Delta_3} \right).
\eeq

\subsection{The effect of Zeeman term on the $b$-Majorana fermions neighboring to vacancies}

When a vacancy is created, the bonds between the vacancy site and the neighboring sites are removed. Therefore, the $b$-Majorana fermions on the sites neighboring the vacancies can no longer form local $Z_2$ fields.

The Zeeman term on the sites neighboring the vacancies can be written in terms of Majorana fermiona as:
\beq
H_{zeeman} = i \frac{h^\alpha}{2} \sum_{j \in \mathcal{N}_\alpha} b_j^\alpha c_j,
\eeq
where $\mathcal{N}$ denotes the set of sites neighboring the vacancies, and the subscript $\alpha$ denotes that the site is originally connected to the vacancy site by a $\alpha$-bond. Physically, this term is interesting, because it describes how the dangling $b$-Majorana, which no longer have to be localized, couple to itinerant $c$-Majorana. This term already indicates that these originally localized dangling $b$-Majorana can become itinerant on the lattice via the Zeeman coupling.

We now show that, through perturbations, Zeeman terms effectively generate the hopping between the dangling $b$-Majorana and $c$-Majorana sitting on the sites further away.

We consider the site $j\in \mathcal{N}$, originally connecting to a vacancy site by a $\alpha$-bond. One can show that a non-trivial perturbation path involves acting Zeeman terms on the two sites connected by a cyan or orange bond successively:
\beq
H_D^{(1)} = -\sum_{j\in \mathcal{N}_\alpha,\langle j,k \rangle} \frac{2h^\beta h^\gamma}{\Delta_{1}} S_j^\beta S_k^\gamma,
\eeq
where the factor $2$ arises from the fact the order of the perturbation can be changed. We do not yet determine which components of the Zeeman coupling are involved. Or explicitly we do not determine $\beta$ and $\gamma$. By using the relation:
\beq
S_j^\beta = \frac{-i}{2} \epsilon_{\alpha,\gamma,\beta} (b_j^\alpha b_j^\gamma),
\eeq
we find:
\beq
S_j^\beta S_k^\gamma = \frac{1}{4} \epsilon_{\alpha,\gamma,\beta} (b_j^\alpha b_j^\gamma) (b_k^\gamma c_k)=\frac{-i}{4} \epsilon_{\alpha,\gamma,\beta} u_{j,k}^\gamma b_j^\alpha c_k.
\eeq
Consequently, the Hamiltonian can be written in terms of Majorana fermions as:
\beq
H_D^{(1)} = \frac{i}{4} \sum_{j\in \mathcal{N}_\alpha,\langle j,k \rangle} \epsilon_{\alpha,\gamma,\beta} \frac{2h^\beta h^\gamma}{\Delta_{in}} u_{j,k}^\gamma b_j^\alpha c_k.
\eeq
This effective Hamiltonian, generated through the perturbation of Zeeman term on the sites in the vicinity of a vacancy, indeed describes the hopping between a dangling $b$-Majorana fermion on the site neighboring to the vacancy and a $c$-Majorana fermion on its neighboring site.

Using the same trick, one can find the next order non-trivial contribution. Due to the geometry of vacancies, the next-nearest-neighboring sites of $j \in \mathcal{N}$ could either be on the boundary of the vacancies (the sites on the blue and orange bonds in Fig.~\ref{figs2}) or not. We need to properly choose the components of Zeeman term to construct a perturbation path. It turns out that the non-trivial contribution in the order $h^3$ is given by:
\begin{align}
H_D^{(2)} = &-\sum_{j\in \mathcal{N}_\alpha} \sum_{\langle j,k,l \rangle} 2(h^\beta)^2 h^\alpha \left( \frac{1}{\Delta_{1} \Delta_{2}} + \frac{1}{\Delta_{1} \Delta_{3}} + \frac{1}{\Delta_{3} \Delta_{2}} \right) S_j^\beta S_k^\alpha S_l^\beta \nonumber \\
& -\sum_{j\in \mathcal{N}_\alpha} \sum_{\langle j,k,m \rangle} 2(h^\beta)^2 h^\alpha \left( \frac{1}{\Delta_{1} \Delta_{2}} + \frac{1}{\Delta_{1} \Delta_{3}} + \frac{1}{\Delta_{3} \Delta_{2}} \right) S_j^\beta S_k^\beta S_m^\alpha,
\end{align}
where $m$ and $l$ label the next-nearest-neighboring sites. Here the site $l$ is on the boundary of the vacancy and the site $m$ is not. In terms of Majorana fermions, the Hamiltonian can be written as:
\beq
H_D^{(2)} = \frac{i}{4} \sum_{j\in \mathcal{N}_\alpha} \sum_{\langle j,k,l \rangle} (h^\beta)^2 h^\alpha \left( \frac{1}{\Delta_{1} \Delta_{2}} + \frac{1}{\Delta_{1} \Delta_{3}} + \frac{1}{\Delta_{3} \Delta_{2}} \right) \left( u_{j,k}^\gamma u_{l,k}^\beta b_j^\alpha c_l - u_{j,k}^\gamma u_{m,k}^\alpha b_j^\alpha c_m \right).
\eeq

\subsection{Effective Hamiltonian accurate to the order of $h^3$}

Eventually, the effective Hamiltonian accurate to the order of $|\bh|^3$ is given by:
\begin{align} \label{Ham_full}
H &= \frac{i}{4}\sum_\alpha \sum_{\langle j,k \rangle} \mathtt{J}_{i,j} c_j c_k + \frac{i}{4} \sum_{\langle j,k,l \rangle} \epsilon_{\alpha,\beta,\gamma} \mathtt{K}_{j,l}^{\alpha,\beta} c_j c_l + \frac{i}{4} \sum_{j \in \mathcal{N}_\alpha} \mathtt{t}_\alpha^{(0)} b_j^\alpha c_j + \frac{i}{4} \sum_{j\in \mathcal{N}_\alpha,\langle j,k \rangle} \epsilon_{\alpha,\gamma,\beta} \mathtt{t}_{\beta,\gamma}^{(1)} b_j^\alpha c_k \nonumber \\
&+ \frac{i}{4} \sum_{j\in \mathcal{N}_\alpha} \sum_{\langle j,k,l \rangle}  \left( \mathtt{t}_{\beta,\alpha;1}^{(2)} b_j^\alpha c_l - \mathtt{t}_{\beta,\alpha;2}^{(2)} b_j^\alpha c_m \right),
\end{align}
where
\beq
\mathtt{t}_\alpha^{(0)}=2h^\alpha,
\eeq
\beq
t_{\beta,\gamma}^{(1)} = \frac{2h^\beta h^\gamma}{\Delta_{in}} u_{j,k}^\gamma,
\eeq
\beq
\mathtt{t}_{\beta,\alpha;1}^{(2)} = (h^\beta)^2 h^\alpha \left( \frac{1}{\Delta_{1} \Delta_{2}} + \frac{1}{\Delta_{1} \Delta_{3}} + \frac{1}{\Delta_{3} \Delta_{2}} \right) u_{j,k}^\gamma u_{l,k}^\beta,
\eeq 
for the site $l$ on the the boundary of the vacancies (forming by the bonds in cyan and orange in Fig.~\ref{figs2}),
\beq
\mathtt{t}_{\beta,\alpha;2}^{(2)} = (h^\beta)^2 h^\alpha \left( \frac{1}{\Delta_{1} \Delta_{2}} + \frac{1}{\Delta_{1} \Delta_{3}} + \frac{1}{\Delta_{3} \Delta_{2}} \right) u_{j,k}^\gamma u_{m,k}^\alpha,
\eeq
for the site $m$ not on the boundary of vacancies, and the effective hopping coefficients $\mathtt{J}_{i,j}$ and $\mathtt{K}_{j,l}^{\alpha,\beta}$ depend on the sites that they connect. 

We first discuss $\mathtt{J}_{i,j}$. If both the two sites $i$ or $j$ are not on the boundary of the vacancies, then $\mathtt{J}_{i,j} = \tilde{J}_\alpha u_{i,j}^{\alpha}$. If the two sites $i$ and $j$ connected by the hopping process denoted by a cyan bond in Fig.~\ref{figs2}, then $\mathtt{J}_{i,j} = \tilde{J}'_\alpha u_{i,j}^{\alpha}$. If the two sites $i$ and $j$ connected by the hopping process denoted by an orange bond in Fig.~\ref{figs2}, then $\mathtt{J}_{i,j} = \tilde{J}''_\alpha u_{i,j}^{\alpha}$.

For the coefficients $\mathtt{K}_{j,l}^{\alpha,\beta}$, they depend on the sites that they connect as follows. If the two sites $j$ or $l$ are not connected by either a red dashed bond or a blue dotted bond in Fig.~\ref{figs2}, $\mathtt{K}_{j,l}^{\alpha,\beta} = \kappa u_{j,k}^\alpha u_{l,k}^\beta$. If the two sites $j$ and $l$ are connected by the hopping process denoted by a red dashed bond, $\mathtt{K}_{j,l}^{\alpha,\beta} = \kappa' u_{j,k}^\alpha u_{l,k}^\beta$. If the two sites $j$ and $l$ are connected by the hopping process denoted by a blue dotted bond, $\mathtt{K}_{j,l}^{\alpha,\beta} = \kappa'' u_{j,k}^\alpha u_{l,k}^\beta$.

When considering a $L\times L$ unit cell lattice with each unit cell labeled by a $2$-dimension index $(n,m)$. We could transform the $2$-dimension indexes into the corresponding $1$-dimension indexes. For the $A$-sublattice at the unit cell $(n,m)$, its $1$-dimension index is given by $nL+m+1$. For the $B$-sublattice in the unit cell $(n,m)$, its $1$-dimension index is given by $nL+m+2$. If there is a pair of vacancies at the $A$-sublattice in the unit cell $(n_A,m_B)$ and the $B$-sublattice in the unit cell $(n_B,m_B)$, their $1$-dimension indexes are $l_A = n_A L+m_A+1$, and $l_B = n_A L+m_A+2$. With this setting, a dangling $b$-Majorana fermion connecting to the $\Lambda=A,B$ vacancy through a $\alpha$-bond could be uniquely labeled as $l_{\Lambda}^{\alpha}$. This labeling strategy could be generalized to the case with arbitrary vacancies.

We define the itinerant Majorana $\psi_{\mathsf{j}}$ by including all the $c$-Majorana fermions $\psi_{\mathsf{j}=j}=c_j$ and the dangling $b$-Majorana fermions $\psi_{\mathsf{j}=j^\alpha}=b_j^\alpha$ adjacent to the vacancies. Then, the Hamiltonian Eq.~(\ref{Ham_full}) is quadratic and can be written in the form: 
\beq \label{effective_Ham}
 H \approx  \frac{i}{4} \sum_{\mathsf{j},\mathsf{k}} \mathcal{T}(\{u_{\langle \mathsf{j},\mathsf{k} \rangle}\}) \psi_{\mathsf{j}} \psi_{\mathsf{k}},
\eeq
where the effective hopping strength $\mathcal{T}(\{u_{\langle \mathsf{j},\mathsf{k} \rangle}\})$ is readable from Eq.~(\ref{Ham_full}). These coefficients depend on the sites that they connect as well as the $Z_2$ gauge fields on the bonds along the closest path connecting them. 

Explicitly, for the $L\times L$ lattice having a pair of vacancies labeled by $l_A$ and $l_B$, we can write the itinerant Majorana as:
\beq \label{Majorana_basis}
\Psi = \left( \psi_1, \psi_2, \cdots, \psi_{l_A-1}, \psi_{l_A+1}, \cdots, \psi_{l_B-1}, \psi_{l_B+1}, \cdots, \psi_{2L^2}, \psi_{l_A^1}, \psi_{l_A^2}, \psi_{l_A^3}, \psi_{l_B^1}, \psi_{l_B^2}, \psi_{l_B^3}  \right),
\eeq
and the effect Hamiltonian can be written as:
\beq
H = \frac{i}{4} \Psi \mathcal{T} \Psi^T.
\eeq

\section{The temporal spin correlation contributed by hybridization of dangling Majorana: Derivation of Eq.~(5)}

The spin correlation is given by:
\beq \label{spin_correlation_vacancies}
\langle S_j^\mu(t) S_l^\nu(0) \rangle = -\frac{1}{4} \frac{\mathrm{Tr} \left[ P e^{-\beta H} e^{iHt} c_j b_j^\mu e^{-iHt} c_l b_l^\nu \right]}{\mathrm{Tr}[Pe^{-\beta H}]}.
\eeq
When $j,l \in \mathcal{N}$, $c_j$, $c_l$, $b_j^\mu$, and $b_l^\nu$ are itinerant Majorana. 

When we have a vector of Majorana, such as Eq.~(\ref{Majorana_basis}), a unitary operator can be expressed as:
\beq
U = e^{\Psi^T h \Psi},
\eeq
with $h$ being a skew-symmetric matrix. One can show the following relation \cite{JozsaS}:
\beq
U^\dag \psi_\mathsf{j} U = \sum_{\nu} [e^{4h}]_{\mathsf{j},\mathsf{k}} \psi_\mathsf{k}.
\eeq
Using this relation, one can prove that Eq.~(\ref{spin_correlation_vacancies}) reduces to:
\beq
\langle S_j^\mu(t) S_l^\nu(0) \rangle = -\frac{1}{4} \sum_{n_1,n_2} [e^{-\mathcal{T}t}]_{n_1,j} [e^{-\mathcal{T}t}]_{n_2,j^\mu} \frac{\mathrm{Tr} \left[ P e^{-\beta H} \psi_{n_1} \psi_{n_2} \psi_l \psi_{l^\nu} \right]}{\mathrm{Tr}[Pe^{-\beta H}]},
\eeq

\subsection{Projecting operator}

Due to the introduction of Majorana, the degrees of the freedom are doubled. We have to introduce a projection operator for local sites:
\beq
P_j = \frac{1+D_j}{2},
\eeq
where $D_j = c_j b_j^x b_j^y b_j^z$. Then the global projection operator is given by:
\beq
P = \left(1 + \prod_j P_j\right)/2.
\eeq

It turns out that for a lattice of a size $L_x \times L_y$ with $L_x \in 2 \mathbb{N}$ and $L_y \in 2\mathbb{N}$, we have:
\beq
\prod_j P_j = (-1)^F (-1)^{F_p},
\eeq
where:
\beq
(-1)^{F_p} = \prod_{j \in \bullet} u_j^x \prod_{j \in \bullet} u_j^y \prod_{j \in \bullet} u_j^z, 
\eeq
and
\beq
(-1)^F = \psi_1 \psi_2 \cdots \psi_{2N+4},
\eeq
with $N=L_1 \times L_2$.

\subsection{Derivation of Eq.~(5) of the main text}

Using the projection operator, the spin correlation can be further simplified as:
\begin{align}
\langle S_j^\mu(t) S_l^\nu(0) \rangle &= -\frac{1}{4} \sum_{n_1,n_2} [e^{-\mathcal{T}t}]_{n_1,j} [e^{-\mathcal{T}t}]_{n_2,j^\mu} \frac{\langle \psi_{n_1}\psi_{n_2} \psi_l \psi_{l^\nu} \rangle_0}{1 + (-1)^{F_p} Z_F/Z_0} \nonumber \\
&-\frac{(-1)^{F_p}}{4} \sum_{n_1,n_2} [e^{-\mathcal{T}t}]_{n_1,j} [e^{-\mathcal{T}t}]_{n_2,j^\mu} \frac{Z_F/Z_0 \langle \psi_{n_1}\psi_{n_2} \psi_l \psi_{l^\nu} \rangle_F}{1 + (-1)^{F_p} Z_F/Z_0},
\end{align}
where:
\beq
Z_0 = \mathrm{Tr}[e^{-\beta H}],~Z_F = \mathrm{Tr}[(-1)^F e^{-\beta H}],
\eeq
and
\beq
\langle \cdots \rangle_0 = \frac{\mathrm{Tr}\left[ e^{-\beta H} \cdots \right]}{\mathrm{Tr}\left[ e^{-\beta H} \right]},~\langle \cdots \rangle_F = \frac{\mathrm{Tr}\left[ (-1)^F e^{-\beta H} \cdots \right]}{\mathrm{Tr}\left[ (-1)^F e^{-\beta H} \right]}.
\eeq

To proceed, we need to use the properties of a skew-symmetric Hermitian matrix $i\mathcal{T}$. Although the derivation can be quite general, we use the case with a pair of vacancies as an explicit example for demonstration. In this situation, $i\mathcal{T}$ has the dimension $2N+4$, and we assume that it can be diagonalized as:
\beq \label{skew}
U^\dag i\mathcal{T} U = \left(\begin{array}{cccccccc} \lambda_1 \\ & \lambda_2 \\
& & \ddots \\ & & & \lambda_{N+2} \\ & & & & -\lambda_1 \\ & & & & & -\lambda_2 \\ & & & & & & \ddots \\ & & & & & & & -\lambda_{N+2} \end{array}\right) = \left(\begin{array}{cccccccc} E_1 \\ & E_2 \\
& & \ddots \\ & & & E_{N+2} \\ & & & & E_{N+3} \\ & & & & & E_{N+4} \\ & & & & & & \ddots \\ & & & & & & & E_{2N+4} \end{array}\right) =D,
\eeq
where:
\beq
U = \left( v_1,v_2,\cdots,v_{N+2},v_{-1},v_{-2},\cdots,v_{-N+2} \right),
\eeq
with $v_{-j}^T v_l = \delta_{j,l}$.

A straightforward calculation shows:
\beq
Z_F/Z_0 = \mathrm{det}[Q] \prod_{n \in \Pi^+} \tanh \frac{\beta \lambda_n}{2},
\eeq
where $\Pi^+$ denotes the non-negative half of the eigenvalue set of $i\mathcal{T}$, and
\beq
Q = \left( v_1,v_{-1},v_2,v_{-2},\cdots,v_{N+2},v_{-N+2} \right).
\eeq
In numerical calculations, $\mathrm{det}[Q]$ can be evaluated more efficiently by using the method in \cite{MasahiroS}.
 
Eventually, using Wick's theorem, the spin correlation can be expressed as:
\begin{align} \label{simplified_spin_correlation}
\langle S_j^\mu(t) S_l^\nu(0) \rangle = &-\frac{1}{4} \sum_{n_1,n_2} [e^{-\mathcal{T}t}]_{n_1,j} [e^{-\mathcal{T}t}]_{n_2,j^\mu} \frac{G_{n_1,n_2}^{(0)} G_{l,l^\nu}^{(0)}-G_{n_1,l}^{(0)} G_{n_2,l^\nu}^{(0)} + G_{n_1,l^\nu}^{(0)} G_{n_2,l}^{(0)}}{1+\eta \prod_{n \in \Pi^+} \tanh \frac{\beta \lambda_n}{2}} \nonumber \\
& -\frac{\eta \prod_{n \in \Pi^+} \tanh \frac{\beta \lambda_n}{2}}{4} \sum_{n_1,n_2} [e^{-\mathcal{T}t}]_{n_1,j} [e^{-\mathcal{T}t}]_{n_2,j^\mu} \frac{G_{n_1,n_2}^{(F)} G_{l,l^\nu}^{(F)}-G_{n_1,l}^{(F)} G_{n_2,l^\nu}^{(F)} + G_{n_1,l^\nu}^{(F)} G_{n_2,l}^{(F)}}{1+\eta \prod_{n \in \Pi^+} \tanh \frac{\beta \lambda_n}{2}},
\end{align}
where $\eta = (-1)^{F_p} \mathrm{det}[Q]$,
\beq
G^{(0)} = \left[ \frac{2}{1+e^{-i\beta \mathcal{T}}} \right],
\eeq
and
\beq
G^{(F)} = \left[ \frac{2}{1-e^{-i\beta \mathcal{T}}} \right].
\eeq

Using Eq.~(\ref{skew}), we can simplify the terms in Eq.~(\ref{simplified_spin_correlation}) one by one. The first term is simplified as:
\beq
[e^{-\mathcal{T}t}]_{n_1,j} [e^{-\mathcal{T}t}]_{n_2,j^\mu} G_{n_1,n_2}^{(0)} G_{l,l^\nu}^{(0)} = [e^{-\mathcal{T}t}]_{j,n_1}^T G_{n_1,n_2}^{(0)} [e^{-\mathcal{T}t}]_{n_2,j^\mu} G_{l,l^\nu}^{(0)} =  [e^{\mathcal{T}t} G^{(0)} e^{-\mathcal{T}t}]_{j,j^\mu} G_{l,l^\nu}^{(0)}.
\eeq
One can show that:
\beq
e^{\mathcal{T}t} G^{(0)} e^{-\mathcal{T}t} = U e^{-iDt} \frac{2}{1+ e^{-\beta D}} e^{iDt} U^\dag.
\eeq
Since $e^{-iDt}$, $\frac{2}{1+ e^{-\beta D}}$ and $e^{iDt}$ are diagonal matrices, their multiplication gives $\frac{2}{1+ e^{-\beta D}}$, which is not time-dependent. Similarly, one can show that $[e^{-\mathcal{T}t}]_{n_1,j} [e^{-\mathcal{T}t}]_{n_2,j^\mu} G_{n_1,n_2}^{(F)} G_{l,l^\nu}^{(F)}$ is also not time dependent. These two terms contribute trivially to $\omega=0$ by Fourier transformation.

The remaining terms can be simplified as:
\beq
[e^{-\mathcal{T}t}]_{n_1,j} [e^{-\mathcal{T}t}]_{n_2,j^\mu} G_{n_1,l}^{(0)} G_{n_2,l^\nu}^{(0)} = [e^{At}]_{j,n_1} G_{n_1,l}^{(0)} [e^{\mathcal{T}t}]_{j^\mu,n_2} G_{n_2,l^\nu}^{(0)} = \left[ e^{-i (i\mathcal{T})t} G^{(0)} \right]_{j,l} \left[ e^{-i (i\mathcal{T})t} G^{(0)} \right]_{j^\mu,l^\nu},
\eeq

\beq
[e^{-\mathcal{T}t}]_{n_1,j} [e^{-\mathcal{T}t}]_{n_2,j^\mu} G_{n_1,l^\nu}^{(0)} G_{n_2,l}^{(0)} = \left[ e^{-i (i\mathcal{T})t} G^{(0)} \right]_{j,l^\nu} \left[ e^{-i (i\mathcal{T})t} G^{(0)} \right]_{j^\mu,l},
\eeq

\beq
[e^{-\mathcal{T}t}]_{n_1,j} [e^{-\mathcal{T}t}]_{n_2,j^\mu} G_{n_1,l}^{(F)} G_{n_2,l^\nu}^{(F)} = \left[ e^{-i (i\mathcal{T})t} G^{(F)} \right]_{j,l} \left[ e^{-i (i\mathcal{T})t} G^{(F)} \right]_{j^\mu,l^\nu},
\eeq

\beq
[e^{-\mathcal{T}t}]_{n_1,j} [e^{-\mathcal{T}t}]_{n_2,j^\mu} G_{n_1,l^\nu}^{(F)} G_{n_2,l}^{(F)} = \left[ e^{-i (i\mathcal{T})t} G^{(F)} \right]_{j,l^\nu} \left[ e^{-i (i\mathcal{T})t} G^{(F)} \right]_{j^\mu,l}.
\eeq
We can rewrite the matrix above in terms of eigenvalues and eigenvectors of $i\mathcal{T}$:
\beq
\left[ e^{-i (i\mathcal{T})t} G^{(0)} \right]_{j,l} \left[ e^{-i (i\mathcal{T})t} G^{(0)} \right]_{j^\mu,l^\nu} = \left[ U \left(e^{-i D t}\right) \left(\frac{2}{1+e^{-\beta D}}\right) U^\dag \right]_{j,l} \left[ U \left(e^{-i D t}\right) \left(\frac{2}{1+e^{-\beta D}}\right) U^\dag \right]_{j^\mu,l^\nu},
\eeq

\beq
\left[ e^{-i (i\mathcal{T})t} G^{(F)} \right]_{j,l} \left[ e^{-i (i\mathcal{T})t} G^{(F)} \right]_{j^\mu,l^\nu} = \left[ U \left(e^{-i D t}\right) \left(\frac{2}{1-e^{-\beta D}}\right) U^\dag \right]_{j,l} \left[ U \left(e^{-i D t}\right) \left(\frac{2}{1-e^{-\beta D}}\right) U^\dag \right]_{j^\mu,l^\nu},
\eeq
 
\beq
\left[ e^{-i (i\mathcal{T})t} G^{(0)} \right]_{j,l^\nu} \left[ e^{-i (i\mathcal{T})t} G^{(0)} \right]_{j^\mu,l} = \left[ U \left(e^{-i D t}\right) \left(\frac{2}{1+e^{-\beta D}}\right) U^\dag \right]_{j,l^\nu} \left[ U \left(e^{-i D t}\right) \left(\frac{2}{1+e^{-\beta D}}\right) U^\dag \right]_{j^\mu,l},
\eeq

\beq
\left[ e^{-i (i\mathcal{T})t} G^{(0)} \right]_{j,l^\nu} \left[ e^{-i (i\mathcal{T})t} G^{(0)} \right]_{j^\mu,l} = \left[ U \left(e^{-i D t}\right) \left(\frac{2}{1-e^{-\beta D}}\right) U^\dag \right]_{j,l^\nu} \left[ U \left(e^{-i D t}\right) \left(\frac{2}{1-e^{-\beta D}}\right) U^\dag \right]_{j^\mu,l}.
\eeq

The matrix elements can be written explicitly as:
\beq
\left[ U \left(e^{-i D t}\right) \left(\frac{2}{1\pm e^{-\beta D}}\right) U^\dag \right]_{j,l} = \sum_{n=1} U(j,n) \frac{2 e^{-iE_n}}{1\pm e^{-\beta E_n}} U^\dag(n,l),
\eeq

\beq
\left[ U \left(e^{-i D t}\right) \left(\frac{2}{1\pm e^{-\beta D}}\right) U^\dag \right]_{j^\mu,l^\nu} = \sum_{n=1}^{N/2} U(j^\mu,n) \frac{2 e^{-iE_n}}{1 \pm e^{-\beta E_n}} U^\dag(n,l^\nu),
\eeq

\beq
\left[ U \left(e^{-i D t}\right) \left(\frac{2}{1\pm e^{-\beta D}}\right) U^\dag \right]_{j,l^\nu} = \sum_{n=1}^{N/2} U(j,n) \frac{2 e^{-iE_n}}{1\pm e^{-\beta E_n}} U^\dag(n,l^\nu),
\eeq

\beq
\left[ U \left(e^{-i D t}\right) \left(\frac{2}{1\pm e^{-\beta D}}\right) U^\dag \right]_{j^\mu,l} = \sum_{n=1}^{N/2} U(j^\mu,n) \frac{2 e^{-iE_n}}{1 \pm e^{-\beta E_n}} U^\dag(n,l),
\eeq
where $U(j,n)$ ($U(j^{\mu},n)$) denotes the $c$-Majorana ($b$-Majorana) component of the eigenfunction corresponding to the energy $E_n$ at the site connecting to the vacancy with index $j$ through a $\mu$-bond. Putting everything together, we find:
\beq \label{spin_correlation_final}
\langle S_j^\mu(t) S_l^\nu(0) \rangle = \frac{-1}{ 4(1+\Gamma)} \sum_{n,m}  e^{-i(E_n+E_m)t} \left(F_0^{\mu,nu}(E_n,E_m)) + F_1^{\mu,nu}(E_n,E_m) \right) + C,
\eeq
where $C$ is the time-independent contribution from the first terms in the first and the second line in Eq.~(\ref{simplified_spin_correlation}), $\Gamma=\eta \prod_{n \in \Pi^+} \tanh \frac{\beta E_n}{2}$,
\beq
F_0^{\mu,\nu}(E_n,E_m) = \frac{\left(U(j,n) U^*(l^\nu,n)U(j^\mu,m) U^*(l,m)-U(j,n) U^*(l,n)U(j^\mu,m) U^*(l^\nu,m)\right)}{(1+e^{-\beta E_n})(1+e^{-\beta E_m})},
\eeq
and
\beq
F_1^{\mu,\nu}(E_n,E_m) = \frac{\Gamma \left(U(j,n) U^*(l^\nu,n)U(j^\mu,m) U^*(l,m)-U(j,n) U^*(l,n) U(j^\mu,m) U^*(l^\nu,m)\right)}{(1-e^{-\beta E_n})(1-e^{-\beta E_m})}
\eeq
We can cast Eq.~(\ref{spin_correlation_final}) into Eq.~(5) of the main text by defining:
\beq
C(\{E_n\},\beta) = \frac{1}{1+\Gamma} = \frac{1}{1+\eta \prod_{n \in \Pi^+} \tanh \frac{\beta \epsilon_n}{2}},
\eeq
\beq
F^{\mu,\nu}(E_n,E_m) = F_0^{\mu,\nu}(E_n,E_m) + F_1^{\mu,\nu}(E_n,E_m).
\eeq

\section{Generality of the $1/T_1$ noise spectrum}

To demonstrate the generality of our results, we present additional results for the $1/T_1$ noise spectrum under the influence of different quantization directions of NV centers, temperatures, and the directions of Zeeman fields. 

\begin{figure}[h!]
  \centering
  \includegraphics[width=0.5\columnwidth]{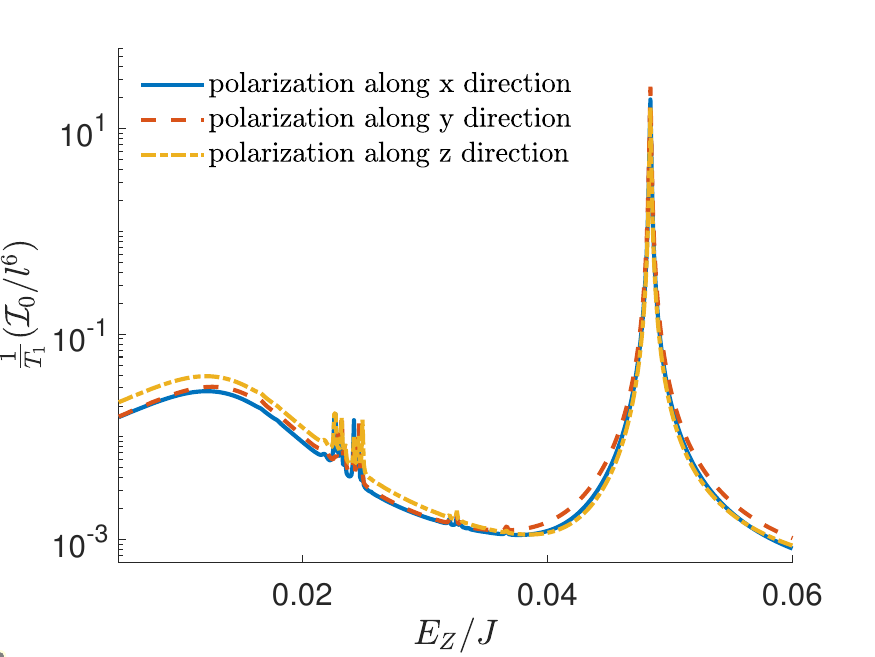}
  \caption{Influence of different quantization directions of NV centers on the $1/T_1$ noise spectrum for the $40 \times 40$ unit cell lattice, with vacancies at the $A$-sublattice in the unit cell $(32,27)$ and the $B$-sublattice in the unit cell $(9,14)$.} 
   \label{S3}
\end{figure}

\begin{figure}[h!]
  \centering
  \includegraphics[width=0.5\columnwidth]{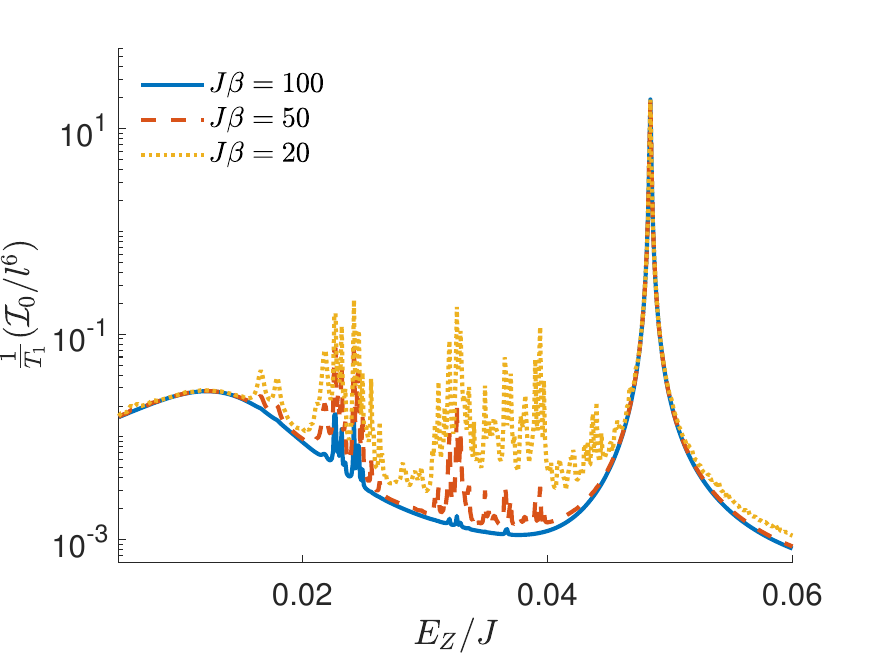}
  \caption{Influence of temperature on the $1/T_1$ noise spectrum for the $40 \times 40$ unit cell lattice, with vacancies at the $A$-sublattice in the unit cell $(32,27)$ and the $B$-sublattice in the unit cell $(9,14)$.} 
   \label{S4}
\end{figure}

\subsection{The influence of NV center polarization}

In Fig.~\ref{S3}, we study how the quantization directions of NV centers affect the $1/T_1$ noise spectrum. Our calculations are performed for a lattice containing $40 \times 40$ unit cells with vacancies at the $A$-sublattice in the unit cell $(32,27)$ and the $B$-sublattice in the unit cell $(9,14)$, which is the same as the example in the main text. We assume that the NV center locates at a height of $d=30a$ and $k_B T=0.01J$. The results shown in Fig.~\ref{S3} indicate that changes in the quantization direction of NV centers have little effect on the $1/T_1$ noise spectrum.

\subsection{The influence of temperature}

In Fig.~\ref{S4}, we plot the $1/T_1$ noise spectrum for three different temperatures by performing calculations on the $40 \times 40$ unit cell lattice with vacancies at the $A$-sublattice of the unit cell $(32,27)$ and the $B$-sublattice of the unit cell $(9,14)$. We assume that the quantization direction of NV centers is along the $x$-direction and that the NV center is placed at a height of $d=30a$ from the lattice plane. From Fig.~\ref{S4}, we find that as the temperature increases, the contributions from the bulk modes become much more significant. However, for $k_B T=0.05J$, which is roughly $T=3$K, the contributions from the bulk modes are still much smaller than those from the hybridized modes and MZMs. The results here indicate that the $1/T_1$ enhancement due to the fractionalized excitations in KQSLs might be sensed at a temperature $\sim 3K$.

\subsection{The influence of the Zeeman field direction}

\begin{figure}[h!]
  \centering
  \includegraphics[width=0.5\columnwidth]{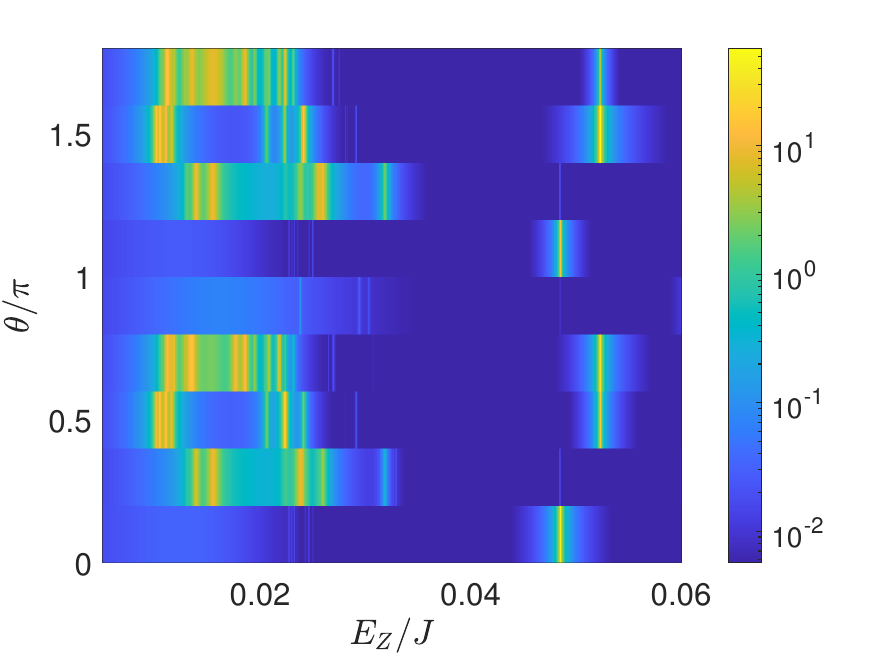}
  \caption{The color map of $1/T_1$ noise spectrum for the in-plane Zeeman field along different directions, which is defined by the angle $\theta$ between the Zeeman field and the $a$-axis.} 
   \label{S5}
\end{figure}

The effective Hamiltonian Eq.~(\ref{effective_Ham}) is obtained by treating the Zeeman term as a perturbation, so we expect that the direction of Zeeman field can affect the hopping matrix and further change the dispersion of the hybridized modes. In this way, the $1/T_1$ noise spectrum can be changed significantly. Here, we focus on in-plane Zeeman field cases, because the out-of-plane $g$-factors of Kitaev materials are very small \cite{KubotaS}.

\begin{figure}[h!]
  \centering
  \includegraphics[width=0.8\columnwidth]{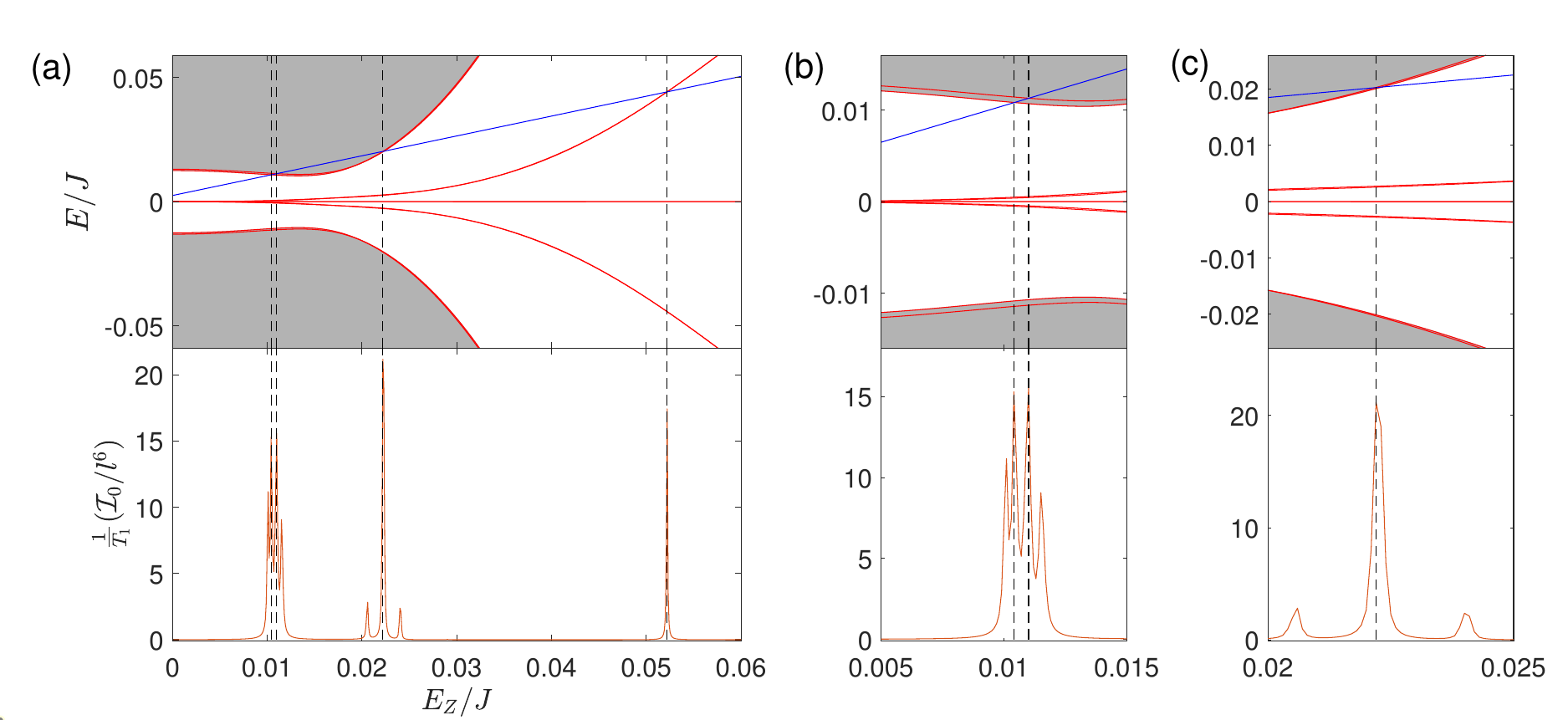}
  \caption{The $1/T_1$ noise spectrum for the case with the in-plane Zeeman field at an angle of $2\pi/5$ from the $a$-axis(as defined in Fig.~1 of the main text). The calculations are performed on the $40 \times 40$ unit cell lattice with vacancies at the $A$-sublattice of the unit cell $(32,27)$ and the $B$-sublattice of the unit cell $(9,14)$: (a) shows the comparison of the dispersion of hybridized modes and the $1/T_1$ spectrum; (b) shows a zoom-in of (a) around $E_Z=0.01J$; (c) shows a zoom-in of (a) around $E_Z=0.02J$. In (b) and (c) we show examples of $1/T_1$ enhancement caused by the hybridized modes locating at the bulk band edges and MZMs. Other visible peaks in the $1/T_1$ spectrum are induced by two hybridized modes whose energy difference matches the working frequency of the NV center.} 
   \label{S6}
\end{figure}

In Fig.~\ref{S5}, we show the $1/T_1$ spectrum for the in-plane Zeeman field along different directions, which are defined by the angle $\theta$ between the Zeeman field and the $a$-axis (defined in Fig.~1 of the main text). We find that the $1/T_1$ noise spectrum exhibits rich properties as the direction of Zeeman field changes. Importantly, the first significant enhancement of $1/T_1$ can occur for $E_z$ as small as $0.01J$. It indicates that, to observe significant $1/T_1$ enhancements due to the fractionalized excitations, the required Zeeman field can be as small as $1/4$ of that with the Zeeman field along the $a$-axis, which can significantly reduce the difficulties for experimental measurements.

To understand the features in Fig.~\ref{S5}, we focus on the case with an in-plane Zeeman field at an angle of $\theta=2\pi/5$ from the $a$-axis and present the $1/T_1$ noise spectrum along with the corresponding dispersion of hybridized modes in Fig.~\ref{S6}. As a result of the change in the Zeeman field direction, $1/T_1$ enhancements can be induced by the hybridized modes locating at the bulk band edges and MZMs, which are shown in Fig.~\ref{S6}(b) and (c). Furthermore, from the $1/T_1$ spectrum shown in Fig.~\ref{S6}(a), in addition to the enhancements induced by the cooperation between hybridized modes and MZMs, we also observe $1/T_1$ enhancements induced by two hybridized modes (peaks other than those coinciding with the vertical black dashed lines), when their energy difference matches the working frequency of the NV center.

\section{Estimation of $1/T_1$ increment due to the fractionalized excitations in site-diluted non-Abelian KQSLs}

In this section, we estimate the $1/T_1$ enhancement by using the data obtained for the example shown in the main text, even though we had shown that the experimental conditions can be further optimized by tuning the Zeeman field direction and the temperatures.

To estimate the enhancement of $1/T_1$, we note that $\Omega_0=3$GHz, which is assumed to be $1/250J$ in the main text. From Fig.~2 of the main text, we find that the working frequency of the NV center, when it matches the energy difference between hybridized modes and MZMs, is $\Omega \sim 30$GHz. Using the gyromagnetic ratio of NV centers, which is $28$GHz/T, we determine that the Zeeman field strength is $\sim 1T$.

The $1/T_1$ is estimated as:
\beq \label{T1_est}
\frac{1}{T_1} = A \frac{g_q^2 g_s^2 \mu_B^4 \mu_0^2}{32 \pi^2 \hbar^2 \Omega d^6},
\eeq
where $A$ is a numerical value. We assume that the $g$-factors of NV centers is $g_q=2$ and the in-plane $g$-factors of Kitaev materials is $g_s=2.5$ \cite{KubotaS}. The parameters in Eq.~(\ref{T1_est}) are given by:
\beq
\mu_B = 9.274 \times 10^{-24}  \mathrm{A \cdot m^2},\quad \mu_0 = 4\pi \times 10^{-7} \mathrm{kg \cdot m / (s^2 \cdot A^2)},\quad \hbar = 1.05 \times 10^{-34} \mathrm{kg \cdot m^2 / s}.
\eeq
Using the data shown in Fig.~3(b) of the main text for $d=30a\sim 10$nm, we find that over one-third area of the supercell has a numerical value larger than $100$. By substituting these values into Eq.~(\ref{T1_est}), we find that over one-third area of the supercell exhibits an enhancement larger than $30$Hz. 

Similarly, from Fig.~3(c) for $d=10a\sim 3.5$nm, we find that over one-third area of the supercell has a numerical value larger than $10000$. Consequently, over one-third area of the supercell exhibits an enhancement larger than $3$kHz.


\begin{thebibliography}{11}

\bibitem{Kitaev}
         A. Kitaev, Anyons in an exactly solved model and beyond,
        \href{https://doi.org/10.1016/j.aop.2005.10.005}{ Ann. Phys. {\bf 321}, 2 (2006)}.
        
\bibitem{Kells}
         G. Kells, J. K. Slingerland, and J. Vala, Description of Kitaev’s honeycomb model with toric-code stabilizers,
        \href{https://doi.org/10.1103/PhysRevB.80.125415}{ Phys. Rev. B {\bf 80}, 125415 (2009)}.
        
\bibitem{Terhal}
         B. Terhal, Quantum error correction for quantum memories,
        \href{https://doi.org/10.1103/RevModPhys.87.307}{ Rev. Mod. Phys. {\bf 87}, 307 (2015)}.
        
\bibitem{Nayak}
         C. Nayak, S. H. Simon, A. Stern, M. Freedman, and S. Das Sarma, Non-abelian anyons and topological quantum computation,
        \href{https://doi.org/10.1103/RevModPhys.80.1083}{ Rev. Mod. Phys. {\bf 80}, 1083 (2008)}.
        
\bibitem{Qi}
         X.-L. Qi, and S.-C. Zhang, Topological insulators and superconductors,
        \href{https://doi.org/10.1103/RevModPhys.83.1057}{ Rev. Mod. Phys. {\bf 83}, 1057 (2008)}.
        
\bibitem{Alicea}
         J. Alicea, New directions in the pursuit of Majorana fermions in solid state systems,
        \href{https://doi.org/10.1088/0034-4885/75/7/076501}{ Rep. Prog. Phys. {\bf 75}, 076501 (2012)}.
        
\bibitem{Sarma}
         S. D. Sarma, M. Freedman, and C. Nayak, Majorana zero
modes and topological quantum computation,
        \href{https://doi.org/10.1038/npjqi.2015.1}{ npj Quantum Information {\bf 1}, 15001 (2015)}.
        
\bibitem{Sato}
         M. Sato, and Y. Ando, Topological superconductors: a review,
        \href{https://doi.org/10.1088/1361-6633/aa6ac7}{ Rep. Prog. Phys. {\bf 80}, 076501 (2017)}.
        
\bibitem{Jackeli}
         G. Jackeli and G. Khaliullin, Mott Insulators in the Strong Spin-Orbit Coupling Limit: From Heisenberg to a Quantum Compass and Kitaev Models,
        \href{https://doi.org/10.1103/PhysRevLett.102.017205}{ Phys. Rev. Lett. {\bf 102}, 017205 (2009)}.
        
\bibitem{Chaloupka}
         J. Chaloupka, G. Jackeli and G. Khaliullin, Kitaev-Heisenberg Model on a Honeycomb Lattice: Possible Exotic Phases in Iridium Oxides A$_2$IrO$_3$,
        \href{https://doi.org/10.1103/PhysRevLett.105.027204}{ Phys. Rev. Lett. {\bf 105}, 027204 (2010)}.
        
\bibitem{Plumb}
         K. W. Plumb, J. P. Clancy, L. J. Sandilands, V. V. Shankar,
Y. F. Hu, K. S. Burch, H.-Y. Kee, and Y.-J. Kim, $\alpha$-RuCl$_3$: A spin-orbit assisted Mott insulator on a honeycomb lattice,
        \href{https://doi.org/10.1103/PhysRevB.90.041112}{ Phys. Rev. B {\bf 90}, 041112(R) (2014)}.
        
\bibitem{Kubota}
         Y. Kubota, H. Tanaka, T. Ono, Y. Narumi, and K. Kindo, Successive magnetic phase transitions in $\alpha$-RuCl$_3$: XY-like frustrated magnet on the honeycomb lattice,
        \href{https://doi.org/10.1103/PhysRevB.91.094422}{ Phys. Rev. B {\bf 91}, 094422 (2015)}.
        
\bibitem{Rousochatzakis}
          I. Rousochatzakis, J. Reuther, R. Thomale, S. Rachel, and
N. B. Perkins, Phase Diagram and Quantum Order by Disorder in the Kitaev $K_1$-$K_2$ Honeycomb Magnet,
        \href{https://doi.org/10.1103/PhysRevX.5.041035}{ Phys. Rev. X {\bf 5}, 041035 (2015)}.
        
\bibitem{Sandilands}
          L. J. Sandilands, Y. Tian, K. W. Plumb, Y.-J. Kim, and
K. S. Burch, Scattering Continuum and Possible Fractionalized Excitations in $\alpha$-RuCl$_3$,
        \href{https://doi.org/10.1103/PhysRevB.91.241110}{ Phys. Rev. B 91, 241110(R) (2015)}.
        
\bibitem{Kim}
          H.-S. Kim, V. Vijay Shankar, A. Catuneanu, and H.-Y. Kee, Kitaev magnetism in honeycomb RuCl$_3$ with intermediate spin-orbit coupling,
        \href{https://doi.org/10.1103/PhysRevB.91.241110}{ Phys. Rev. B 91, 241110(R) (2015)}.
        
\bibitem{Kim16}
          H.-S. Kim, and H.-Y. Kee, Crystal structure and magnetism in $\alpha$-RuCl$_3$: An ab initio study,
        \href{https://doi.org/10.1103/PhysRevB.93.155143}{ Phys. Rev. B 93, 155143 (2016)}.
        
\bibitem{Nasu}
          J. Nasu, J. Knolle, D. L. Kovrizhin, Y. Motome, and R. Moessner. Fermionic response from fractionalization in an insulating two-dimensional magnet,
        \href{https://doi.org/10.1038/nphys3809}{ Nat. Phys. {\bf 12}, 912 (2016)}.  
        
\bibitem{Arnab}
          A. Banerjee, C. A. Bridges, J.-Q. Yan, A. A. Aczel, L. Li,
M. B. Stone, G. E. Granroth, M. D. Lumsden, Y. Yiu, J.
Knolle, S. Bhattacharjee, D. L. Kovrizhin, R. Moessner,
D. A. Tennant, D. G. Mandrus, and S. E. Nagler Proximate Kitaev quantum spin liquid behaviour in a honeycomb magnet,
        \href{https://doi.org/10.1038/nmat4604}{ Nat. Mater. {\bf 15}, 733 (2016)}.  
     
\bibitem{Arnab17}
          A. Banerjee, J. Yan, J. Knolle, C. A. Bridges, M. B. Stone, M. D. Lumsden, D. G. Mandrus, D. A. Tennant, R. Moessner, S. E. Nagler Neutron scattering in the proximate quantum spin liquid $\alpha$-RuCl$_3$,
        \href{https://doi.org/10.1126/science.aah6015}{ Science {\bf 356}, 1055 (2017)}.    
        
\bibitem{Trebst}
          S. Trebst, Kitaev Materials,
        \href{https://doi.org/10.48550/arXiv.1701.07056}{arXiv:1701.07056 (2017)}.
        
\bibitem{Takagi}
          H. Takagi, T. Takayama, G. Jackeli, G. Khaliullin, and
S. E. Nagler, Concept and realization of Kitaev quantum spin liquids,
        \href{https://doi.org/10.1038/s42254-019-0038-2}{Nat. Rev. Phys. {\bf 1}, 264 (2019)}.
        
\bibitem{Hermanns}
          M. Hermanns, I. Kimchi, and J. Knolle, Physics of the Kitaev model: fractionalization, dynamical correlations, and material connections,
        \href{https://doi.org/10.1146/annurev-conmatphys-033117-053934}{Annu. Rev. Condens. Matter Phys. 9, 17 (2018)}.
        
\bibitem{Baek}
         S.-H. Baek, S.-H. Do, K.-Y. Choi, Y. S. Kwon, A. U. B.
Wolter, S. Nishimoto, J. van den Brink, and B. Buchner, Evidence for a Field-Induced Quantum Spin Liquid in $\alpha$-RuCl$_3$,
        \href{https://doi.org/10.1103/PhysRevLett.119.037201}{Phys. Rev. Lett. {\bf 119}, 037201 (2017)}.
        
\bibitem{Sears}
         J. A. Sears, Y. Zhao, Z. Xu, J. W. Lynn, and Y.-J. Kim, Phase diagram of $\alpha$-RuCl$_3$ in an in-plane magnetic field,
        \href{https://doi.org/10.1103/PhysRevB.95.180411}{Phys. Rev. B {\bf 95}, 180411(R) (2017)}.
        
\bibitem{Wolter}
         A. U. B. Wolter, L. T. Corredor, L. Janssen, K. Nenkov,
S. Schönecker, S.-H. Do, K.-Y. Choi, R. Albrecht, J.
Hunger, T. Doert, M. Vojta, and B. Buchner, Field-induced quantum criticality in the Kitaev system $\alpha$-RuCl$_3$,
        \href{https://doi.org/10.1103/PhysRevB.96.041405}{Phys. Rev. B {\bf 96}, 041405(R) (2017)}.
        
\bibitem{Leahy}
         I. A. Leahy, C. A. Pocs, P. E. Siegfried, D. Graf, S.-H.
Do, K.-Y. Choi, B. Normand, and M. Lee, Anomalous Thermal Conductivity and Magnetic Torque Response in the Honeycomb Magnet $\alpha$-RuCl$_3$,
        \href{https://doi.org/10.1103/PhysRevLett.118.187203}{Phys. Rev.
Lett. {\bf 118}, 187203 (2017)}.

\bibitem{Jansa}
         N. Jansa, A. Zorko, M. Gomilsek, M. Pregelj, K. W. Kramer, D. Biner, A. Biffin, C. Ruegg, and M. Klanjsek, Observation
of two types of fractional excitation in the Kitaev honeycomb
magnet,
        \href{https://doi.org/10.1038/s41567-018-0129-5}{Nature Physics {\bf 14}, 786 (2018)}.

\bibitem{Arnab18}
         A. Banerjee, P. Lampen-Kelley, J. Knolle, C. Balz, A. A. Aczel, B. Winn, Y. Liu, D. Pajerowski, J. Yan, C. A. Bridges, A. T. Savici, B. C. Chakoumakos, M. D. Lumsden, D. A. Tennant, R. Moessner, D. G. Mandrus,
and S. E. Nagler, Excitations in the field-induced quantum spin liquid state of $\alpha$-RuCl$_3$,
        \href{https://doi.org/10.1038/s41535-018-0079-2}{npj Quant. Mater. {\bf 3}, 8 (2018)}.
        
\bibitem{Hentrich}
         R. Hentrich, A. U. B. Wolter, X. Zotos, W. Brenig, D. Nowak, A. Isaeva, T. Doert, A. Banerjee, P. LampenKelley, D. G. Mandrus, S. E. Nagler, J. Sears, Y.-J. Kim, B. Buchner, and C. Hess, Unusual Phonon Heat Transport in $\alpha$-RuCl$_3$: Strong Spin-Phonon Scattering and Field-Induced Spin Gap,
        \href{https://doi.org/10.1103/PhysRevLett.120.117204}{Phys. Rev. Lett. {\bf 120}, 117204 (2018)}.
        
\bibitem{Widmann}
         S. Widmann, V. Tsurkan, D. A. Prishchenko, V. G. Mazurenko, A. A. Tsirlin, and A. Loidl, Thermodynamic evidence of fractionalized excitations in $\alpha$-RuCl$_3$,
        \href{https://doi.org/10.1103/PhysRevB.99.094415}{Phys. Rev. B {\bf 99}, 094415 (2019)}.
        
\bibitem{Balz}
         C. Balz, P. Lampen-Kelley, A. Banerjee, J. Yan, Z. Lu, X. Hu, S. M. Yadav, Y. Takano, Y. Liu, D. A. Tennant, M. D. Lumsden, D. Mandrus, and S. E. Nagler, Finite field regime for a quantum spin liquid in $\alpha$-RuCl$_3$,
        \href{https://doi.org/10.1103/PhysRevB.100.060405}{Phys. Rev. B {\bf 100}, 060405(R) (2019)}.
        
\bibitem{Czajka}
         P. Czajka, T. Gao, M. Hirschberger, P. Lampen-Kelley, A. Banerjee, J. Yan, D. G. Mandrus, S. E. Nagler, and N. P. Ong, Oscillations of the thermal conductivity in the spin-liquid state of $\alpha$-Rucl$_3$,
        \href{https://doi.org/10.1038/s41567-021-01243-x}{Nat. Phys. {\bf 17}, 915 (2021)}.
        
\bibitem{Tanaka}
         O. Tanaka, Y. Mizukami, R. Harasawa, K. Hashimoto, N. Kurita, H. Tanaka, S. Fujimoto, Y. Matsuda, E.-G. Moon, and T. Shibauchi, Thermodynamic evidence for a field-angle-dependent Majorana gap in a Kitaev spin liquid,
        \href{https://doi.org/10.1038/s41567-021-01488-6}{Nat. Phys. {\bf 18}, 429 (2022)}.
       
\bibitem{Czajka23}
         P. Czajka, T. Gao, M. Hirschberger, P. Lampen-Kelley, A. Banerjee, N. Quirk, D. G. Mandrus, S. E. Nagler, and N. P. Ong, Planar thermal hall effect of topological bosons in the kitaev magnet $\alpha$-RuCl$_3$,
        \href{https://doi.org/10.1038/s41563-022-01397-w}{Nat. Mater. {\bf 22}, 36 (2023)}. 

\bibitem{Kasahara}
         Y. Kasahara, T. Ohnishi, Y. Mizukami, O. Tanaka, S. Ma, K. Sugii, N. Kurita, H. Tanaka, J. Nasu, Y. Motome, T. Shibauchi, and Y. Matsuda, Majorana quantization and half-integer thermal quantum hall effect in a kitaev spin liquid,
        \href{https://doi.org/10.1038/s41586-018-0274-0}{Nature {\bf 559}, 227 (2018)}.
        
\bibitem{Yamashita}
         M. Yamashita, J. Gouchi, Y. Uwatoko, N. Kurita, and H. Tanaka, Sample dependence of half-integer quantized thermal Hall effect in the Kitaev spin-liquid candidate $\alpha$-RuCl$_3$,
        \href{https://doi.org/10.1103/PhysRevB.102.220404}{Phys. Rev. B {\bf 102}, 220404(R) (2020)}.
        
\bibitem{Yokoi}
         T. Yokoi, S. Ma, Y. Kasahara, S. Kasahara, T. Shibauchi, N. Kurita, H. Tanaka, J. Nasu, Y. Motome, C. Hickey, S. Trebst, and Y. Matsuda, Half-integer quantized anomalous thermal Hall effect in the Kitaev material candidate $\alpha$-RuCl$_3$,
        \href{https://doi.org/10.1126/science.aay5551}{Science {\bf 373}, 568 (2021)}.
        
\bibitem{Bruin}
         J. A. N. Bruin, R. R. Claus, Y. Matsumoto, N. Kurita, H. Tanaka, and H. Takagi, Robustness of the thermal Hall effect close to half-quantization in $\alpha$-RuCl$_3$,
        \href{https://doi.org/10.1038/s41567-021-01501-y}{Nat. Phys. {\bf 18}, 401 (2022)}.
        
\bibitem{Sahasrabudhe}
         A. Sahasrabudhe, D. A. S. Kaib, S. Reschke, R. German, T. C. Koethe, J. Buhot, D. Kamenskyi, C. Hickey, P. Becker, V. Tsurkan, A. Loidl, S. H. Do, K. Y. Choi, M. Gruninger, S. M. Winter, Z. Wang, R. Valenti, and P. H. M. van Loosdrecht, High-field quantum disordered state in $\alpha$-RuCl$_3$: Spin flips, bound states, and multiparticle continuum,
        \href{https://doi.org/10.1103/PhysRevB.101.140410}{Phys. Rev. B {\bf 101}, 140410(R) (2020)}.
        
\bibitem{Bachus}
         S. Bachus, D. A. S. Kaib, Y. Tokiwa, A. Jesche, V. Tsurkan, A. Loidl, S. M. Winter, A. A. Tsirlin, R. Valenti, and P. Gegenwart, Thermodynamic Perspective on Field-Induced Behavior of $\alpha$-RuCl$_3$,
        \href{https://doi.org/10.1103/PhysRevLett.125.097203}{	Phys. Rev. Lett. {\bf 125}, 097203 (2020)}.
        
\bibitem{Aasen}
         D. Aasen, R. S. K. Mong, B. M. Hunt, D. Mandrus, and
J. Alicea, Electrical Probes of the Non-Abelian Spin Liquid in Kitaev Materials,
        \href{https://doi.org/10.1103/PhysRevX.10.031014}{Phys. Rev. X {\bf 10}, 031014 (2020)}.
        
\bibitem{Konig}
         E. J. Konig, M. T. Randeria, and B. Jack, Tunneling spec- ¨
troscopy of quantum spin liquids,
        \href{https://doi.org/10.1103/PhysRevLett.125.267206}{Phys. Rev. Lett. {\bf 125}, 267206 (2020)}.
        
\bibitem{Pereira}
         R. G. Pereira and R. Egger, Electrical access to ising anyons in
kitaev spin liquids,
        \href{https://doi.org/10.1103/PhysRevLett.125.227202}{Phys. Rev. Lett. {\bf 125}, 227202 (2020)}.
        
\bibitem{Feldmeier}
         J. Feldmeier, W. Natori, M. Knap, and J. Knolle, Local probes
for charge-neutral edge states in two-dimensional quantum magnets,
        \href{https://doi.org/10.1103/PhysRevB.102.134423}{Phys. Rev. B {\bf 102}, 134423 (2020)}.
        
\bibitem{Udagawa}
         M. Udagawa, S. Takayoshi, and T. Oka, Scanning tunneling microscopy as a single majorana detector of kitaev’s chiral spin liquid,
        \href{https://doi.org/10.1103/PhysRevLett.126.127201}{Phys. Rev. Lett. {\bf 126}, 127201 (2021)}.
        
\bibitem{Bauer}
         T. Bauer, L. R. D. Freitas, R. G. Pereira, and R. Egger, Scanning
tunneling spectroscopy of majorana zero modes in a kitaev spin liquid,
        \href{https://doi.org/10.1103/PhysRevB.107.054432}{Phys. Rev. B {\bf 107}, 054432 (2023)}.
        
\bibitem{Takahashi}
         M. O. Takahashi, M. G. Yamada, M. Udagawa, T. Mizushima, and S. Fujimoto, Non-local spin correlation as a signature of ising anyons trapped in vacancies of the kitaev spin liquid,
        \href{https://doi.org/10.1103/PhysRevLett.131.236701}{Phys. Rev. Lett. {\bf 131}, 236701 (2023)}.
        
\bibitem{Kao}
         W.-H. Kao, N. B. Perkins, G. B Halasz, Vacancy spectroscopy of non-Abelian Kitaev spin liquids,
        \href{https://doi.org/10.1103/PhysRevLett.132.136503}{Phys. Rev. Lett. {\bf 132}, 136503 (2024)}.
        
\bibitem{Halasz}
         G. B. Halasz, Gate-Controlled Anyon Generation and Detection in Kitaev Spin Liquids,
        \href{https://doi.org/10.1103/PhysRevLett.132.206501}{Phys. Rev. Lett. {\bf 132}, 206501 (2024)}.
        
\bibitem{Go}
         A. Go, J. Jung, E.-G. Moon, Vestiges of Topological Phase Transitions in Kitaev Quantum Spin Liquids,
        \href{https://doi.org/10.1103/PhysRevLett.122.147203}{Phys. Rev. Lett. {\bf 122}, 147203 (2019)}.
        
\bibitem{Choi}
         W. Choi, K. H. Lee, and Y. B. Kim, Vestiges of Topological Phase Transitions in Kitaev Quantum Spin Liquids,
        \href{https://doi.org/10.1103/PhysRevLett.124.117205}{Phys. Rev. Lett. {\bf 124}, 117205 (2020)}.
        
\bibitem{Willans}
         A. J. Willans, J. T. Chalker, and R. Moessner, Disorder in a quantum spin liquid: Flux binding and local moment formation, 
        \href{https://doi.org/10.1103/PhysRevLett.104.237203}{Phys. Rev. Lett. {\bf 104}, 237203 (2010)}.
        
\bibitem{Willans11}
         A. J. Willans, J. T. Chalker, and R. Moessner, Site dilution in the kitaev honeycomb model, 
        \href{https://doi.org/10.1103/PhysRevB.84.115146}{Phys. Rev. B {\bf 84}, 115146 (2011)}.
        
\bibitem{Kao20}
         W.-H. Kao, J. Knolle, G. B. Halasz, R. Moessner, and N. B. Perkins, Vacancy-Induced Low-Energy Density of States in the Kitaev Spin Liquid, 
        \href{https://doi.org/10.1103/PhysRevX.11.011034}{Phys. Rev. X {\bf 11}, 011034 (2021)}.
        
\bibitem{Imamura24}
         K. Imamura, Y. Mizukami, O. Tanaka, R. Grasset, M. Konczykowski, N. Kurita, H. Tanaka, Y. Matsuda, M. G. Yamada, K. Hashimoto, T. Shibauchi, Defect-Induced Low-Energy Majorana Excitations in the Kitaev Magnet $\alpha$-RuCl$_3$, 
        \href{https://doi.org/10.1103/PhysRevX.14.011045}{Phys. Rev. X {\bf 14}, 011045 (2024)}.
        
\bibitem{Casola}
         F. Casola, T. van der Sar, and A. Yacoby, Probing condensed matter physics with magnetometry based on nitrogen-vacancy centres in diamond,
        \href{https://doi.org/10.1038/natrevmats.2017.88}{Nat. Rev. Mater. {\bf 3}, 17088 (2018)}.
        
\bibitem{Agarwal}
         K. Agarwal, R. Schmidt, B. Halperin, V. Oganesyan, G. Zarand, M. D. Lukin, and E. Demler, Magnetic noise spectroscopy as a probe of local electronic correlations in two-dimensional systems,
        \href{https://doi.org/10.1103/PhysRevB.95.155107}{Phys. Rev. B {\bf 95}, 155107 (2017)}.
        
\bibitem{Ariyaratne}
         A. Ariyaratne, D. Bluvstein, B. A. Myers, and A. C. B. Jayich, Nanoscale electrical conductivity imaging using a nitrogen-vacancy center in diamond,
        \href{https://doi.org/10.1038/s41467-018-04798-1}{Nat. Commun. {\bf 9}, 2406 (2018)}.
        
\bibitem{Dolgirev}
         P. E. Dolgirev, S. Chatterjee, I. Esterlis, A. A. Zibrov, M. D. Lukin, N. Y. Yao, and E. Demler, Characterizing two-dimensional superconductivity via nanoscale noise magnetometry with single-spin qubits,
        \href{https://doi.org/10.1103/PhysRevB.105.024507}{Phys. Rev. B {\bf 105}, 024507 (2022)}.
        
\bibitem{Khoo}
         J. Y. Khoo, F. Pientka, P. A. Lee, and I. S. Villadiego, Probing the quantum noise of the spinon Fermi surface with NV centers,
        \href{https://doi.org/10.1103/PhysRevB.106.115108}{Phys. Rev. B {\bf 106}, 115108 (2022)}.
        
\bibitem{Lee}
         P. A. Lee and S. Morampudi, Proposal to detect emergent gauge field and its Meissner effect in spin liquids using NV centers,
        \href{https://doi.org/10.1103/PhysRevB.107.195102}{Phys. Rev. B {\bf 107}, 195102 (2023)}.
        
\bibitem{Chatterjee}
         S. Chatterjee, J. F. Rodriguez-Nieva, and E. Demler, Diagnosing phases of magnetic insulators via noise magnetometry with spin qubits,
        \href{https://doi.org/10.1103/PhysRevB.99.104425}{Phys. Rev. B {\bf 99}, 104425 (2019)}.
        
\bibitem{Nieva}
         J. F. Rodriguez-Nieva, D. Podolsky, and E. Demler, Probing hydrodynamic sound modes in magnon fluids using spin magnetometers,
        \href{https://doi.org/10.1103/PhysRevB.105.174412}{Phys. Rev. B {\bf 105}, 174412 (2022)}.
        
\bibitem{Langsjoen}
         L. S. Langsjoen, A. Poudel, M. G. Vavilov, and R. Joynt, Qubit relaxation from evanescent-wave Johnson noise,
        \href{https://doi.org/10.1103/PhysRevA.86.010301}{Phys. Rev. A {\bf 86}, 010301(R) (2012)}.
 
\bibitem{support}
         See supplementary materials for details.
        
\bibitem{Baskaran}
         G. Baskaran, S. Mandal,and R. Shankar, Exact Results for Spin Dynamics and Fractionalization in the Kitaev Model, 
        \href{https://doi.org/10.1103/PhysRevLett.98.247201}{Phys. Rev. Lett. {\bf 98}, 247201 (2007)}.
        
\bibitem{Knolle}
         J. Knolle, D. L. Kovrizhin, J. T. Chalker, and R. Moessner, Dynamics of a Two-Dimensional Quantum Spin Liquid: Signatures of Emergent Majorana Fermions and Fluxes, 
        \href{https://doi.org/10.1103/PhysRevLett.112.207203}{Phys. Rev. Lett. {\bf 112}, 207203 (2014)}.

\bibitem{Eichstaedt}
         C. Eichstaedt, Y. Zhang, P. Laurell, S. Okamoto, A. G. Eguiluz, and T. Berlijn, Deriving models for the Kitaev spin-liquid candidate material $\alpha$-RuCl$_3$ from first principles,
        \href{https://doi.org/10.1103/PhysRevB.100.075110}{Phys. Rev. B~{\bf 100}, 075110 (2019)}.
        
\bibitem{Myers}
         B. A. Myers, A. Ariyaratne, and A. C. B. Jayich, Double-Quantum Spin-Relaxation Limits to Coherence of Near-Surface Nitrogen-Vacancy Centers,
        \href{https://doi.org/10.1103/PhysRevLett.118.197201}{Phys. Rev. Lett.~{\bf 118}, 197201 (2017)}.
        
\bibitem{Sangtawesin}
         S. Sangtawesin, B. L. Dwyer, S. Srinivasan, J. J. Allred, L. V. H. Rodgers, K. De Greve, A. Stacey, N. Dontschuk, K. M. O'Donnell, D. Hu, D. A. Evans, C. Jaye, D. A. Fischer, M. L. Markham, D. J. Twitchen, H. Park, M. D. Lukin, N. P. de Leon, Origins of diamond surface noise probed by correlating single spin measurements with surface spectroscopy,
        \href{https://doi.org/10.1103/PhysRevX.9.031052}{Phys. Rev. X~{\bf 9}, 031052 (2019)}.
        
\bibitem{Jarmola}
         A. Jarmola, V. M. Acosta, K. Jensen, S. Chemerisov, and D. Budker, Temperature- and Magnetic-Field-Dependent Longitudinal Spin Relaxation in Nitrogen-Vacancy Ensembles in Diamond,
        \href{https://doi.org/10.1103/PhysRevLett.108.197601}{Phys. Rev. Lett.~{\bf 108}, 197601 (2012)}.
        
\bibitem{Roccapriore}
         K. M. Roccapriore, M. G. Boebinger, O. Dyck, A. Ghosh, R. R. Unocic, S. V. Kalinin, and M. Ziatdinov, Probing Electron Beam Induced Transformations on a Single-Defect Level via Automated Scanning Transmission Electron Microscopy,
        \href{https://doi.org/10.1021/acsnano.2c07451}{ACS Nano~{\bf 16}, 17116 (2022)}.
        
\bibitem{Thomas}
         J. C. Thomas, W. Chen, Y. Xiong, B. A. Barker, J. Zhou, W. Chen, A. Rossi, N. Kelly, Z. Yu, D. Zhou, S. Kumari, E. S. Barnard, J. A. Robinson, M. Terrones, A. Schwartzberg, D. F. Ogletree, E. Rotenberg, M. M. Noack, S. Griffin, A. Raja, D. A. Strubbe, G.-M. Rignanese, A. Weber-Bargioni, G. Hautier, A substitutional quantum defect in WS$_2$ discovered by high-throughput computational screening and fabricated by site-selective STM manipulation,
        \href{https://doi.org/10.1038/s41467-024-47876-3}{Nat. Commun. {\bf 15}, 3556 (2024)}.
        
\bibitem{Zhou}
         B. B. Zhou, P. C. Jerger, K.-H. Lee, M. Fukami, F. Mujid, J. Park, and D. D. Awschalom, Spatiotemporal Mapping of a Photocurrent Vortex in Monolayer MoS$_2$ Using Diamond Quantum Sensors,
        \href{https://doi.org/10.1103/PhysRevX.10.011003}{Phys. Rev. X~{\bf 10}, 011003 (2020)}.
        
\bibitem{Zhang}
         X.-Y. Zhang, Y.-X. Wang, T. A. Tartaglia, T. Ding, M. J. Gray, K. S. Burch, F. Fafti, and B. B. Zhou, AC Susceptometry of 2D van der Waals Magnets Enabled by the Coherent Control of Quantum Sensors,
        \href{https://doi.org/10.1103/PRXQuantum.2.030352}{PRX Quantum~{\bf 2}, 030352 (2021)}.
        
\bibitem{Kumar}
         J. Kumar, D. Yudilevich, A. Smooha, I. Zohar, A. K. Pariari, R. Stohr, A. Denisenko, M. Hucker, and A. Finkler, Room Temperature Relaxometry of Single Nitrogen Vacancy Centers in Proximity to $\alpha$-RuCl$_3$ Nanoflakes,
        \href{https://doi.org/10.1021/acs.nanolett.3c05090}{Nano. Lett.~{\bf 24}, 4793 (2024)}.
        
\bibitem{Welter}
        P. Welter, J. Rhensius, A. Morales, M. S. Wörnle, C.-H. Lambert, G. Puebla-Hellmann, P. Gambardella, and C. L. Degen, Scanning nitrogen-vacancy center magnetometry in large in-plane magnetic fields,
        \href{https://doi.org/10.1063/5.0084910}{Appl. Phys. Lett. {\bf 120}, 074003 (2022)}.
        
\bibitem{Beaver}
        N. M. Beaver, N. Voce, P. Meisenheimer, R. Ramesh, and P. Stevenson, Optimizing Off-Axis Fields for Two-Axis Magnetometry with Point Defects,
        \href{https://doi.org/10.1063/5.0214004}{Appl. Phys. Lett. {\bf 124}, 254001 (2024)}.
        
\bibitem{Stepanov}
        V. Stepanov, F. H. Cho, C. Abeywardana, and S. Takahashi, High-frequency and high-field optically detected magnetic resonance of nitrogen-vacancy centers in diamond,
        \href{https://doi.org/10.1063/1.4908528}{Appl. Phys. Lett. {\bf 106}, 063111 (2015)}.
        
\bibitem{Fortman}
        B. Fortman, L. Mugica-Sanchez, N. Tischler, C. Selco, Y. Hang, K. Holczer, and S. Takahashi, Electron-electron double resonance detected NMR spectroscopy using ensemble NV centers at 230 GHz and 8.3 Tesla,
        \href{https://doi.org/10.1063/5.0055642}{J. Appl. Phys. {\bf 130}, 083901 (2021)}.
        
\bibitem{Ren}
        Y. Ren, C. Selco, D. Kawashiri, M. Coumans, B. Fortman, L. S. Bouchard, K. Holczer, S. Takahashi, Demonstration of NV-detected $^{13}$C NMR at 4.2T,
        \href{https://doi.org/10.1103/PhysRevB.108.045421}{Phys. Rev. B {\bf 108}, 045421 (2023)}.
        
\bibitem{Takahashib}
        M. O. Takahashi, W.-H. Kao, S. Fujimoto, and N. B. Perkins, Z$_2$ flux binding to higher-spin impurities in the Kitaev spin liquid: mechanisms and implications,
        \href{https://doi.org/10.48550/arXiv.2409.02190}{	arXiv:2409.02190 (2024)}.

\end{thebibliography}

\begin{thebibliography}{70}

\bibitem{KitaevS}
         A. Kitaev, Anyons in an exactly solved model and beyond,
        \href{https://doi.org/10.1016/j.aop.2005.10.005}{ Ann. Phys. {\bf 321}, 2 (2006)}.
        
\bibitem{JozsaS}
         R. Jozsa, and A. Miyake, Matchgates and classical simulation of
quantum circuits,
        \href{https://doi.org/10.1098/rspa.2008.0189}{ Proc. R. Soc. A {\bf 464}, 3089 (2008)}.
        
\bibitem{MasahiroS}
         M. O. Takahashi, M. G. Yamada, M. Udagawa, T. Mizushima, and S. Fujimoto, Non-local spin correlation as a signature of ising anyons trapped in vacancies of the kitaev spin liquid,
        \href{https://doi.org/10.1103/PhysRevLett.131.236701}{Phys. Rev. Lett. {\bf 131}, 236701 (2023)}.
        
\bibitem{KubotaS}
         Y. Kubota, H. Tanaka, T. Ono, Y. Narumi, and K. Kindo, Successive magnetic phase transitions in $\alpha$-RuCl$_3$: XY-like frustrated magnet on the honeycomb lattice,
        \href{https://doi.org/10.1103/PhysRevB.91.094422}{ Phys. Rev. B {\bf 91}, 094422 (2015)}.

\end{thebibliography}
\end{document}